\documentclass[%
 reprint,
 amsmath,amssymb,
pra,superscriptaddress
]{revtex4-2}
\usepackage{graphicx}
\usepackage{graphics}
\usepackage{dcolumn}
\usepackage{bm}
\usepackage{amsfonts}       
\usepackage{nicefrac}       
\usepackage{microtype}      
\usepackage{lipsum}
\usepackage{subcaption}
\usepackage{amsmath}
\usepackage{xcolor}
\def\ba{{\bf a}}
\def\bb{{\bf b}}

\def\b1{{1\!\!1}}

\def\cH{{\ca H}}

\def\bC{{\mathbb C}}           

\def\bE{{\mathbb E}}

\def\bN{{\mathbb N}}

\def\bR{{\mathbb R}}

\def\beq{\begin{eqnarray}}
\def\eeq{\end{eqnarray}}

\newcommand{\ca}[1]{{\cal #1}}         

\newtheorem{theorem}{\em Theorem}[section]

\newtheorem{remark}[theorem]{\em Remark}

\begin{document}

\preprint{arXiv:2003.09961}

\title{Bell inequality violation by entangled single photon states generated from a laser, a LED, or a halogen lamp}

\author{Matteo Pasini}
\altaffiliation[Now at ]{Hanson Lab, QuTech, Delft University of Technology, Holland.}
\affiliation{%
 Nanoscience Laboratory, Department of Physics, University of Trento, Italy\\
}%
\author{Nicolò Leone}%
\affiliation{%
 Nanoscience Laboratory, Department of Physics, University of Trento, Italy\\
}%

\author{Sonia Mazzucchi}
\affiliation{%
 Department of Mathematics and TIFPA-INFN, University of Trento, Italy\\
}%

\author{Valter Moretti}
\affiliation{%
 Department of Mathematics and TIFPA-INFN, University of Trento, Italy\\
}%

\author{Davide Pastorello}%
\affiliation{%
 Department of Mathematics and TIFPA-INFN, University of Trento, Italy\\
}%
\affiliation{%
Department of Information Engineering and Computer Science, University of Trento, Italy\\
}%

\author{Lorenzo Pavesi}%
\email[Correpondence should be addressed to: ]{lorenzo.pavesi@unitn.it}
\affiliation{%
 Nanoscience Laboratory, Department of Physics, University of Trento, Italy\\
}%

\date{\today} 

\begin{abstract}
In single-particle or intraparticle entanglement, two degrees of freedom of a single 
particle, e.g., momentum and  polarization of a single photon, are entangled. Single-particle entanglement (SPE) is a resource that can be exploited both in quantum communication protocols and in experimental tests of noncontextuality based on the Kochen-Specker theorem. Here, we show that single-particle entangled states of single photons can be produced from attenuated classical sources of light. To experimentally certify the single-particle entanglement, we perform a Bell test, observing a violation of the Clauser, Horne, Shimony and Holt (CHSH) inequality. We show that single-particle entanglement can be achieved even in a classical light beam, provided that first-order coherence is maintained between the degrees of freedom involved in the entanglement. This demonstrates that cheap, compact, and low power photon sources can be used to generate single-particle entangled photons which could be a resource for quantum technology applications.
\end{abstract}

\maketitle
\section{Introduction}
Entanglement is one of the most popular and controversial  phenomena in quantum mechanics. Originally introduced as a source of paradoxes, such as the Schr\"{o}dinger’s cat and the Einstein, Podolsky, Rosen phenomenology \cite{EPR1935}, nowadays it has become a  fundamental resource in the blooming areas of quantum information and quantum computing \cite{nielsen2002quantum,Macchiavello04,Bennett92,Ekert91,Ekert98}.
As clearly shown in the pioneering work by J. Bell \cite{bell1964}, entanglement produces correlations between the outcomes of measurements that do not have a classical counterpart. In other words, they cannot be explained in terms of a local realistic hidden variable theory.
From a mathematical point of view, the notion of entanglement relies on the tensor product structure of the Hilbert space associated to a quantum system with (at least) two independent degrees of freedom (DoF). In the simple case of a bipartite system, entanglement (or non-separability) of a state is defined in terms of its Schmidt rank \cite{nielsen2002quantum}.
From a physical point of view, it is possible to distinguish between two kinds of entanglement. Entanglement between the DoF of distinct particles, e.g. photons \cite{Kwiat95}, is called {\em interparticle entanglement}. On the contrary, entanglement of the DoF of a single particle is called {\em single-particle entanglement} (SPE) or {\em intraparticle entanglement} \cite{Can05,Michler00,Gadway09,Markiewicz2019}. Even if SPE does not produce the non-local correlations typical of interparticle entangled systems (Einstein’s spooky action at distance), it is still a signature of quantum behaviour. Indeed, SPE has been used  in tests of noncontextuality related to Kochen-Specker theorem and Hardy paradox \cite{Michler00,simon2000,huang2003,hasegawa2004,hasegawa2011,xin2012,karimi2014hardy}. In particular, violation of Bell-type inequalities for SPE systems proves the impossibility of describing  quantum phenomenology in terms of a realistic non contextual hidden variable theory\footnote{Experimental  tests for non contextuality based on Bell inequality violation can be classified as  “statistical tests”. Indeed,   they rely  on  the contrast between   the probabilistic  predictions of quantum mechanics   and  the corresponding predictions of  non contextual hidden-variables theories (NCHVT), according to which Bell inequality  cannot be violated.
On the other hand a set of experimental test of noncontextuality have been proposed, which in the spirit of Kochen-Specker theorem are “non statistical” and are  usually called “all-or-nothing tests”. They are based on the contradiction between the prediction of quantum mechanics and NCHVT in the case of suitable couples of compatible observables are jointly measured. In particular, the experimental test of ref. \cite{simon2000,huang2003} checks whether the measurement outcomes of two specific observables have always opposite (as predicted by quantum mechanics) or equal  values (as predicted by NCHVT). This task can be accomplished by looking at the count rate of eight different detectors. If quantum mechanics is right only four of them will register photons and the others won’t register anything, while this situation would be completely reversed in the case of validity of NCHVT. For a more detailed discussion of this topic we refer to \cite{simon2000}.}. From  the technological side, SPE states find applications in recent quantum information protocols \cite{lee2003quantum,Massa18,del2018two,Adhikari15,Beige02,saha2016robustness}.

The generation of entangled photons pairs is typically based on a high power laser that pumps a non-linear crystal, where two entangled photons are produced by either Spontaneous Parametric Down Conversion \cite{Magnitskiy15} or by Four Wave Mixing \cite{Takesue04}. In the case of SPE cheaper and simpler light sources might be used
since correlated photons pairs are no longer necessary. However, single-particle entangled states have been only generated starting from heralded single photons \cite{Gadway09,valles2014generation}. These require high power laser, hence are expensive and with a low generation efficiency.

In this paper, we show that under suitable experimental conditions SPE can also be obtained from classical and cheap attenuated light sources. This result is certified by a Bell test, observing a violation of the Clauser, Horne, Shimony and Holt (CHSH) inequality\cite{CHSH1969}, using several types of sources: a laser, a LED and a halogen lamp.  We argue that in all three cases the results of our experiment reveal the quantum nature of the phenomenon. This fact may appear quite surprising at a first glance, as the SPE states of the electromagnetic field have {\em positive $P$-function}, so that they enjoy  classical features according to the Glauber-Sudarshan's approach to quantum optics.\cite{mandel1995optical} In fact,  as we shall see in this work, the quantum nature shows up only when measuring the single photon properties of a light beam.

In addition, in this paper we remark the difference between our results and experiments on the so called \emph{classical entanglement}, where correlations between DoF of a classical beam of light lead to  violation of Bell-like inequalities that can be observed by measuring light intensities \cite{spreeuw1998,Spreeuw01,borges2010,Kagalwala2013,Aiello2015}. In fact, in this case, the CHSH inequality violation can be fully predicted by the   classical theory of electromagnetic fields and it is just a witness of the nonseparability of the DoF of the light beam as well as its first order coherence properties. This similarity  between SPE and {classical entanglement} could induce to conclude that only interparticle entanglement is truly quantum because SPE can be simulated by means of classical fields.

Following this reasoning the notion of \emph{non-locality} can be elected as the basic property differing quantum and classical entanglement \cite{Spreeuw01,karimi2015classical}. However, this viewpoint is challenged by other recent interpretations \cite{Khrennikov2020,korolkova2019} supporting the idea that a distinguishing feature of quantum theory is contextuality which can be observed even in absence of  non-locality. In fact, classical entanglement differs fundamentally from entanglement between DoF of a single quantum system as SPE is. Indeed, if one measures intensities of signals in different measurement channels instead of counting discrete events (clicks of detectors), then CHSH violation provides information only on {\em collective properties} of the light beam. On the other hand, in our case, the use of suitably attenuated sources and single photon avalanche detectors (SPAD) allow to obtain information on properties of the single photons composing the light beam. In fact, what matters here is not the cumulative statistics, but the whole time ordered sequence of outcomes of measurements of couples of single-particle observables. In this case, when CHSH inequality is violated by the statistics of these single-particle observable outcomes, the observed statistics cannot be explained in terms of a non contextual realistic hidden variable theory of the single photons in the beam.  This opens the doors to applications to quantum information \cite{Adhikari15,pironio2010random} since each photon can become the carrier of a couple of entangled qubits.  

Eventually, it is worth discussing the connection with the results presented in \cite{reid1986violations,zukowski2016bell}, where the authors propose alternative Bell-type inequalities for quantum optical fields which use intensities as observables. These results are rather different from ours since they essentially refer to intense beams consisting of couples of  (interparticle) entangled photons, but it is important to mention them since they are able to exhibit  a quantum signature even in apparently classical-like contexts.

This paper is organized as follow. In section \ref{secBell}, a theoretical analysis of the experiment for the Bell tests of SPE based on CHSH inequality is presented and some technical notions are introduced and discussed. In Section \ref{Experiment}, the experimental set-up to generate SPE and the measurements of CHSH violations for the different light sources are presented. In section \ref{discussion}, a comparison between SPE, interparticle entanglement and the so-called classical entanglement is discussed and an interpretation of the obtained results for their possible technological impact is reported. Section \ref{conclusions} concludes the paper with a summary of the main results. In Appendix, there is the analysis of the coherence length/time that is crucial to motivate the finite-dimensional description of the considered SPE states. This analysis also includes a description of the effective $S$-parameter.

\section{Theoretical analysis of the experiment}\label{secBell}

\begin{figure*}
	\includegraphics[width = \linewidth]{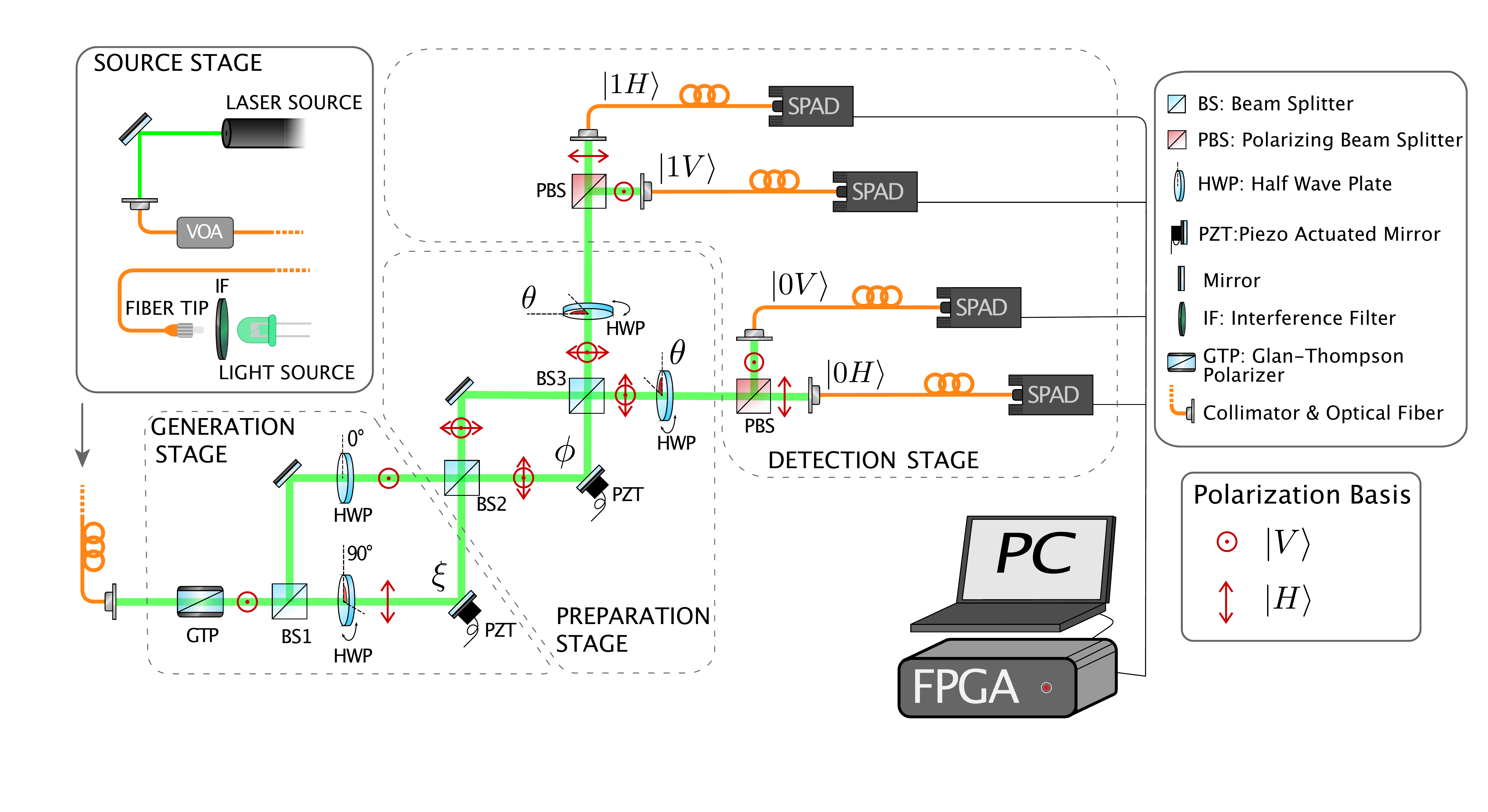}
	\caption{Schematic of the setup to generate and test single-particle entanglement. The description of the different components and symbols is given in the text.}
	\label{fig:setup}
\end{figure*}

We set-up an experiment to generate and verify SPE (fig. \ref{fig:setup}). As an entanglement witness, the set-up aims at implement a test of single-particle entanglement by using the version of the Bell inequality \cite{bell1964} due to Clauser, Horne, Shimony and Holt (CHSH) \cite{CHSH1969} as in  \cite{Michler00, Barreiro05}. For SPE, we use the momentum and the polarization DoF of photons.
For momentum, we  fix two possible different wave vectors $\textbf{k}_0, \textbf{k}_1$, i.e.,  two directions of propagation in the experimental setup, together with a common 
frequency  $\nu = \frac{c}{2\pi}|\textbf{k}_0|=  \frac{c}{2\pi}|\textbf{k}_1|$. Hence, in this setting, the  Hilbert space describing the momentum DoF reduces to   a 2-dimensional space $\mathcal{H}_M$ spanned by the associated qubit basis states  $|0\rangle, |1\rangle$. For polarization, we  take the vertical and horizontal directions (with respect to the propagation plane), defining the basis $|V\rangle, |H\rangle$  of the polarization Hilbert space $\mathcal{H}_P$. The space of our two-qubit composite system is the four-dimensional $\mathcal{H}_S = \mathcal{H}_M \otimes \mathcal{H}_P$ spanned by 
\begin{equation}
\{|0V\rangle, |0H\rangle, |1V\rangle, |1H\rangle\},
\label{eq:4statebasis}
\end{equation}
where henceforth  $|{XY}\rangle =|X\rangle\otimes |Y\rangle$.

Checking the validity of CHSH inequality traditionally requires the measurement of several pairs of mutually compatible observables (one in $\mathcal{H}_M$ and the other in $\mathcal{H}_P$) on a fixed entangled state. As illustrated below, this is physically  (unitarily) equivalent to keep fixed a couple of observables and to change the measured state accordingly (see Eq. (\ref{eqidprop}) below).  This second possibility is  experimentally more convenient and justifies  the setup of our experiment.
 
The experimental setup can be functionally divided in three stages, as illustrated in Fig. \ref{fig:setup}: (I) the generation, (II) the preparation,  and (III) the detection stages. Let us briefly explain these stages.

\begin{figure*}
\includegraphics[width = \linewidth]{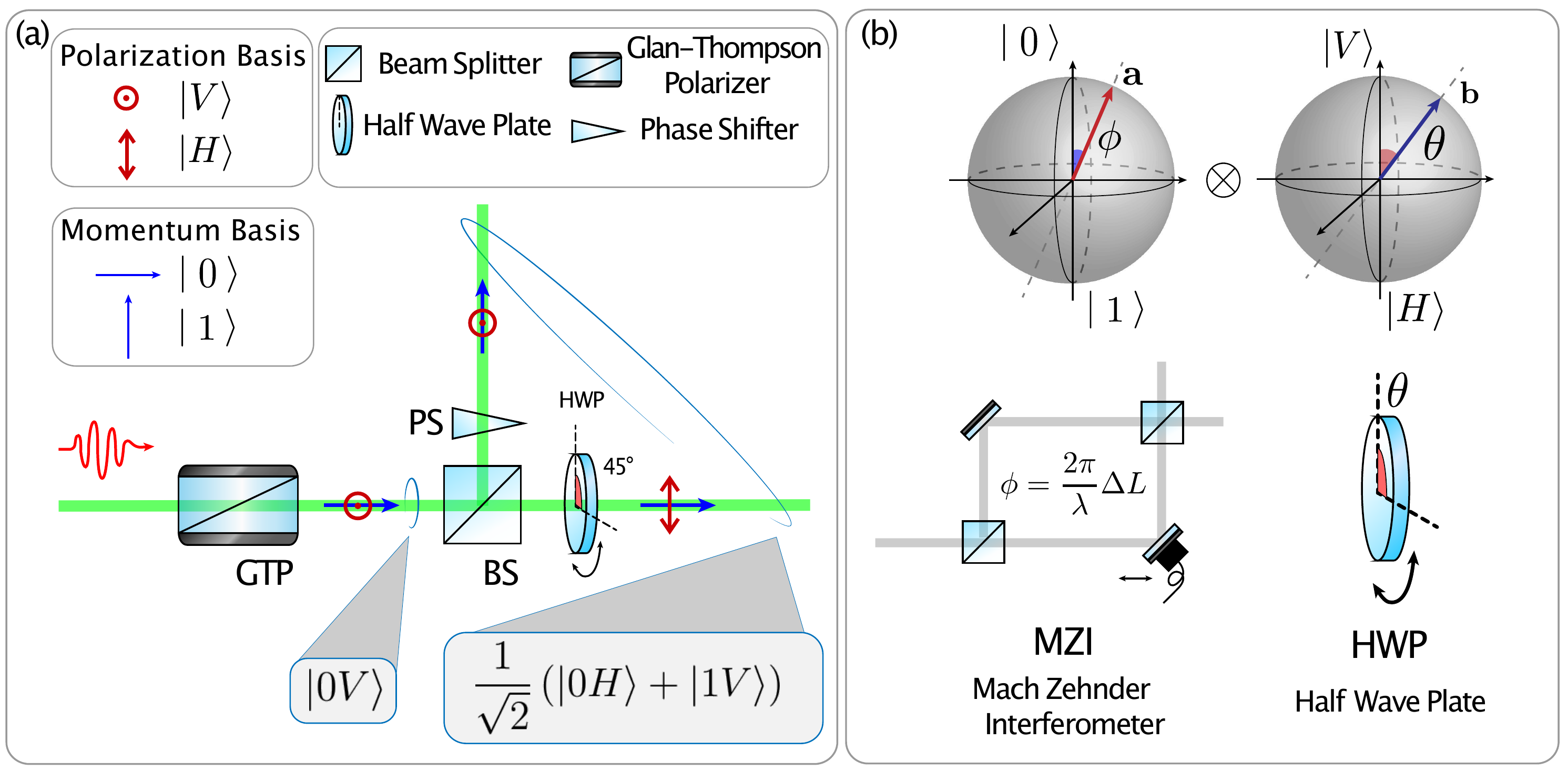}
\caption{(a) Setup for the generation of single-particle entangled states. (b) Momentum and polarization bases in the Bloch sphere and the optical elements constituting the preparation stage of Fig. 1.}
\label{fig:intraparticle}
\end{figure*}

{\bf (I)}  In the first generation stage, the single-photon entangled state is  generated. For the photon entering the setup, the input polarization and momentum  are defined by
using a Glan Thompson Polarizer and a collimator  so that the state of the photon is $|\psi\rangle = |0V\rangle$. Next, 
we  generate an  entangled state  from the initial state $|\psi\rangle$, as schematically shown in \ref{fig:intraparticle}(a). A beam splitter (BS) puts the initial state in a superposition of momentum states $|\psi\rangle = (|0\rangle+|1\rangle)\otimes |V\rangle$ and a half wave plate (HWP) rotates the polarization of  photons  with  momentum ${\bf k}_1$ so that the state entering the second beam splitter is the Bell state  $|\Psi_+\rangle$ \cite{Braunstein1992}

\begin{equation}
|\Psi_+\rangle = \frac{1}{\sqrt{2}}(|0H\rangle + |1V\rangle).
\label{eq:entangledstate}
\end{equation}
A more realistic and detailed description of the actual state of the photons is given in Appendix \ref{SEC1}. \\
{\bf (II)} The following preparation stage performs the rotation $U_\ba\otimes U_\bb$ in the  space $\mathcal{H}_M \otimes \mathcal{H}_P$ transforming the entangled state $|\Psi_+\rangle$ 
to a prepared state $|\psi'_{\textbf{a},\textbf{b}}\rangle$. 
This can be done in  different ways 
depending on a couple  $(\textbf{a},\textbf{b})$ of unit vectors in the Bloch sphere (Fig. \ref{fig:intraparticle}(b)). All transformations are performed by
suitable unitary maps $U_\ba$ and $U_\bb$ in the qubit spaces
(see (\ref{UUU})-(\ref{UUUU}) in Appendix).
The Mach Zehnder interferometer (MZI) in the second part of the setup acts as a momentum-qubit gate $U_\ba$, rotating the state by an angle $\phi = \frac{2\pi \Delta L}{\lambda}$, where $\Delta L$ is the path difference 
in the two arms and $\lambda$ the photon wavelength \cite{Gadway09}. The angle $\phi$ determines the  vector $\textbf{a}$  in the Bloch sphere (see Fig. \ref{fig:intraparticle}(b)) associated to  the 1-particle  observable $ O^{\bf a}={\bf a}\cdot \boldsymbol{\sigma}$ related to the momentum DoF, where $\boldsymbol{\sigma} =(\boldsymbol{\sigma}_x, \boldsymbol{\sigma}_y, \boldsymbol{\sigma}_z)$ are the associated Pauli matrices. In particular, the orthogonal projectors $\{P^\ba_{x}\}_{x=\pm1}$ associated to the eigenvectors of $O^{\bf a}$ are obtained by applying the unitary map $U_\ba$ to the projectors $\{P^M_{+1}=|0\rangle \langle 0|, P^M_{-1}=|1\rangle \langle 1|\}$ , associated to the standard basis of $\mathcal{H}_M $, i.e. 
\begin{equation}\label{Pba} P^\ba_{x}=U_\ba ^\dag P^M_xU_\ba, \qquad x=\pm 1.\end{equation}

Two half wave plates (HWP), one in each output port of the MZI (fig. \ref{fig:setup}), 
with the fast axis rotated by the same amount $\vartheta$, perform a rotation in the polarization space 
by an angle $\theta = 2\vartheta$ with respect to the vertical direction. This transformation in the qubit space $\mathcal{H}_P $ can be described in terms of a unitary map $U_\bb$. The  angle $\theta $ determines the vector
$\textbf{b}$  in the Bloch sphere, i.e. the 1-particle observable  $ O^{\bf b}={\bf b}\cdot{\bf \boldsymbol{\sigma}}$ related to the polarization DoF. As above, the orthogonal projectors $\{ P^\bb_{y} \}_{y=\pm 1}$ associated to the eigenvectors of $O^{\bf b}$ can be obtained as 
\begin{equation}\label{Pbb} P^\bb_{y}=U_\bb ^\dag P^P_y U_\bb, \qquad y=\pm 1,\end{equation} where $\{P^P_{+1}=|H\rangle \langle H|, P^P_{-1}=|1\rangle \langle 1|\}$ are the projectors associated to the basis $\{|H\rangle, |V\rangle\}$ of $\mathcal{H}_P $.

{\bf (III)}   In the third detection stage, the joint measurement of the observables $O^{\bf a}\otimes I$ and $ I\otimes O^{\bf b}$ is performed on the state $|\Psi_+\rangle $, the probability of obtaining a particular pair $(x,y)$ of outcomes being given by $\{\langle \Psi_+|P^\ba_x\otimes P^\bb_y|\Psi_+\rangle\}_{x,y=\pm 1}$.  By \eqref{Pba} and \eqref{Pbb} this is unitarily equivalent to the measurement of  momentum and polarization of the prepared states $|\psi'_{\textbf{a},\textbf{b}}\rangle$, i.e. to computing their components $\{\langle \psi'_{\textbf{a},\textbf{b}}|P^M_x\otimes P^P_y|\psi'_{\textbf{a},\textbf{b}}\rangle\}_{x,y=\pm 1}$ with respect to  the elements of the standard base (\ref{eq:4statebasis}): 
\begin{equation}
\label{eqidprop}\langle \Psi_+|P^\ba_x\otimes P^\bb_y|\Psi_+\rangle = \langle \psi'_{\textbf{a},\textbf{b}}|P^M_x\otimes P^P_y|\psi'_{\textbf{a},\textbf{b}}\rangle\:.\end{equation}
From the experimental point of view,  momentum  is measured by  looking at the two output arms of the preparation stage, while polarization is measured by using Polarizing Beam Splitters (PBS) to spatially separate the polarization components. Four collimators, one for each state, couple the photons into optical 
fibers connected to four Silicon SPADs. These identify the four measurement channels.

\subsection{States}\label{sec1}
 
In general, the state of the electromagnetic field entering the first beam splitter of Fig.\ref{fig:setup} is a multi-particle state. In the experiments, we considered two types of states. The first is a coherent superposition of pure states of finite number of particles in the same mode 
\begin{equation}\label{state-c}|\Psi \rangle := \sum_{n=0}^{+\infty} C_n |n_\psi \rangle\:,\quad \mbox{where} \quad \sum_{n=0}^{+\infty} |C_n|^2 =1\end{equation}
	and it is typically obtained   by a short time laser pulse.  The second is an incoherent superposition of pure states of finite number of particles in the same mode
	\begin{equation}\label{state-i}\rho  := \sum_{n=0}^{+\infty} P_n |n_\psi\rangle\langle n_\psi| \:,\quad \mbox{where} \quad \sum_{n=0}^{+\infty} P_n=1\end{equation}
	and it is typically obtained after frequency filtration of an  incoherent 
source (LED/halogen lamp ) or a laser beam \cite{Wiseman2016}.

In \eqref{state-c} and \eqref{state-i}, $|n_{\psi}\rangle$ refers to the one-particle input state $|\psi\rangle$ selected by mode filtration. In particular, $|\psi\rangle = |0V\rangle$. This notation is the standard of 2nd quantization in the Fock space 
$|n_\psi \rangle = \frac{1}{\sqrt{n!}} (a_\psi^\dagger)^n |vac\rangle$,
where $|vac\rangle$ is the  vacuum state. \\
In the case of an ideal laser, the representation \eqref{state-c} is valid, far above threshold and at atomic physics time-scale, with
$C_n = e^{-\mu/2} \frac{\mu^{n/2}}{\sqrt{n!}}$,  where $\mu = \langle N\rangle$ is equal to the expected value of the number operator $N$ and the standard deviation is given by  $\sigma_n = \sqrt{\langle N\rangle}$. However, the phase between states with different number of 
particles quickly becomes undefined, through a process of phase
diffusion \cite{Wiseman2016} and the emitted state settles in the form \eqref{state-i} with a Poissonian distribution 
$P_n = e^{-\mu}\frac{\mu^n}{n!}$ so that $\langle N\rangle = \mu$ and $\sigma_n = \sqrt{\langle N\rangle}$ is still valid. In the case of a mode filtered thermal light at temperature $T$, the form 
\eqref{state-i} of the input state  is valid with
$P_n = \frac{1}{1+\langle n\rangle}\left( \frac{\langle n\rangle}{\langle n\rangle +1}\right)^n\:,$ where
$\langle n\rangle = \frac{1}{\exp\{\hbar \omega/k_B T\}-1}$ and $\omega =c |\bf{k}|$.   This model is also valid for a LED source.

\subsection{Action on multiparticle states by the setup}\label{Secmanyparticles}

Since all the optical devices used  to manipulate photons in the first two stages of the setup are linear, their  action on photons can be described by unitary operators $R_0,R$, acting in the  one-particle space $\mathcal{H}_M\otimes \mathcal{H}_P$.
In particular, generation of single photon entangled states only needs linear optical elements that act separately on the two DoF (see Fig. \ref{fig:intraparticle}a). 
The net effect of the generation stage on single photons is to produce  one-particle entangled states:
$$|\psi\rangle  \to  R_0|\psi \rangle := |\Psi_+\rangle$$
where, as said  $|\psi\rangle =|0V\rangle$ and $|\Psi_+\rangle$ is the Bell state \eqref{eq:entangledstate}. 
The cumulative effect of the generation and of the preparation stages  is similarly described as
$$|\psi \rangle \to |\psi'\rangle =R |\psi\rangle\:.$$

Actually $R$ and, therefore, $|\psi'\rangle$  depend on the choice 
of the parameters ${\bf a}, {\bf b}$  used to prepare the state in the second stage of the setup (and for that reason we shall use later the more precise notation
$|\psi'_{{\bf a}, {\bf b}}\rangle$).
In view of the said linearity, the action on the actual  multi-particle states handled by the setup is represented by the unitary operator $U_R$ acting in the Fock space and completely defined by the requirements
$$U_R a_\psi^\dagger U_R^\dagger = a^\dagger_{\psi'}\:, \quad U_R|vac\rangle = |vac\rangle\:, \quad |\psi'\rangle = R|\psi\rangle\:.$$
Therefore, the net action on the  states \eqref{state-c}  and \eqref{state-i} is:
  \begin{eqnarray*} 
	&&|\Psi \rangle \mapsto |\Psi' \rangle := U_R |\Psi \rangle = \sum_{n=0}^{+\infty} C_n |n_{\psi'} \rangle\:,\\
	&&\rho  \mapsto \rho' := U_R \rho U^\dagger_R = \sum_{n=0}^{+\infty} P_n |n_{\psi'}\rangle\langle n_{\psi'}| \:.\end{eqnarray*}

\subsection{Measured tests}
The final stage of the experiment consists of the measurement of a set of observables $Q$ attaining only values $0$ or $1$. For each of these tests $Q$, the probability to be found true (outcome $1$) in  the transformed state is:
\begin{itemize}
	\item[(1)]  For a coherent superposition of pure states of finite number of particles, the probability that $Q$ is true is
	$$\langle \Psi' |Q|\Psi' \rangle := \sum_{n,m=0}^{+\infty} \overline{C_{m}}C_{n}\langle m_{\psi'}|Q|n_{\psi'}\rangle\:.$$
	
	\item[(2)]  For an incoherent superposition of pure states of finite number of particles, the probability that $Q$ is true is
	$$tr(\rho'Q)  := \sum_{n=0}^{+\infty} P_n \langle n_{\psi'}|Q|n_{\psi'}\rangle\:.$$
\end{itemize}
The crucial observation is that {\em each used test $Q$ commutes with the observable number of particle}.
As a consequence, in the case (1), $$\mbox{$\langle m_{\psi'}|Q|n_{\psi'}\rangle=0$ if $m \neq n$}\quad$$ so that
$$\quad \langle \Psi' |Q|\Psi' \rangle := \sum_{n=0}^{+\infty} |C_{n}|^2\langle n_{\psi'}|Q|n_{\psi'}\rangle\:.$$
Hence, the statistics of the final measurement cannot distinguish between the case of an initial state given in terms of a coherent superposition of pure states with given number of particles $|\Psi \rangle := \sum_{n=0}^{+\infty} C_n |n_\psi \rangle\:$, or a corresponding incoherent superposition $\rho  := \sum_{n=0}^{+\infty} |C_n|^2 |n_\psi\rangle\langle n_\psi| \:$.
Therefore, all the analysis can be  performed referring to pure states with fixed number of particles and, finally, combining them with the corresponding weights $P_n$ or $|C_n|^2$ according to the case.

Since, according to the above discussion,  we can separately consider states with fixed number of particles, let us assume 
that the Fock state $|n_\psi\rangle$ enters the first stage of the setup.
What we actually measure on {\em each} photon in the prepared state  $|n_{\psi'}\rangle$  (actually $|n_{\psi'_{{\bf a},{\bf b}}}\rangle$) are the values of two observables, each with two possible outcomes or, equivalently, an observable with  {\em four} possible outcomes  henceforth denoted   
by $1,2,3,4$ for notation simplicity. 
In fact, for $n=1$ (i.e. one single photon) the four tests $Q_1,Q_2,Q_3,Q_4$ are one-to-one associated with the four measurement channels of the setup and correspond to the orthogonal projectors 
associated to the basis (\ref{eq:4statebasis}), i.e., 
$|0V\rangle\langle 0V|, 
|1V\rangle\langle 1V|,|0H\rangle\langle 0H|,|1H\rangle\langle 1H|$.
These tests  are {\em mutually compatible}, {\em pairwise exclusive}, and {\em exhaustive}:
$$[Q_i,Q_j]=0\:,  \quad Q_iQ_j =0 \quad \mbox{for $i\neq j$}, $$ $$\quad Q_1+Q_2+Q_3+Q_4 =I\:.$$

In the general case of the $n$-particle subspace of the Fock space,  the single-particle tests  $Q_1,Q_2,Q_3,Q_4$ induce  a class of tests
$Q_{(n_1, n_2, n_3, n_4)}$,
where $(n_1,n_2,n_3, n_4)$ range in the set of  strings  of $4$ natural numbers (including $0$)  
such that $n_1+n_2+n_3+n_4 = n$.  

By definition,  $Q_{(n_1, n_2, n_3, n_4)}$ is validated if and only if $n_1$ particles produce the outcome $1$,  
$n_2$ particles produce the outcome $2$, $n_3$ particles produce the outcome $3$, and $n_4$ particles produce the outcome $4$.
These multi-particle tests are 
{\em mutually compatible}, 
$$[Q_{(n_1,n_2,n_3, n_4)},Q_{(m_1,m_2,m_3, m_4)}]=0,$$
{\em pairwise exclusive} 
$$ Q_{(n_1,n_2,n_3, n_4)}Q_{(m_1,m_2,m_3, m_4)} =0 \quad$$ 
for $(n_1,n_2,n_3, n_4)\neq (m_1,m_2,m_3, m_4)$,
and {\em exhaustive}
$$ \sum_{n_1+n_2+n_3+n_4 = n} Q_{(n_1,n_2,n_3, n_4)}=I\:.$$
These tests  are exhaustive, 
they account for all possible outcomes of the detection stage of the setup when the input state includes $n$ indistinguishable particles. In terms of one-particle tests $Q_j$, these new operators  are defined as 
\begin{multline}\label{n-particle case0} Q_{(n_1,n_2,n_3,n_4)} \equiv\\ \sum_{\pi }Q_1^{\pi(1)}\cdots Q_1^{\pi(n_1)}Q_2^{\pi(n_1+1)}
\cdots Q_2^{\pi(n_1+n_2)} Q_3^{\pi(n_1+n_2+1)}\cdots \\\cdots Q_3^{\pi(n_1+n_2+n_3)} Q_4^{\pi(n_1+n_2+n_3+1)}\cdots Q_4^{\pi(n_1+n_2+n_3+n_4)} \end{multline}
where the sum  is taken over all permutations $\pi$ of the set of indexes $\{1,\ldots,n\}$
and the notation $A_1^{j_1}\cdots A_n^{j_n}$, with $\{j_1,...,j_n\}=\{1,...,n\} $, stands for the operator acting on the  tensor 
product of $n$ copies of the Hilbert space $\mathcal{H}_S$, and  $A_l^{J_l}$ indicates  the operator
$$ A_l^{J_l}\equiv \underbrace{I\otimes \cdots \otimes I }_{j_l-1\, \hbox{\scriptsize factors $I$}}\otimes \underbrace{A_l}_ {j_l\, \hbox{\scriptsize th position}} \otimes\underbrace{  I\otimes\cdots \otimes I}_{n-j_l \, \hbox{\scriptsize    factors $I$}}\:.$$
For the state $|n_{\psi'}\rangle$, the probability to find $n_1$ particles with outcome $1$, $n_2$ particles with outcome $2$, $n_3$ 
particles with outcome $3$,  and $n_4$ particles with outcome $4$ (where  $n_1+n_2+n_3+n_4 = n$) can be computed from (\ref{n-particle case0}) to be
\begin{multline}\label{distribution}\langle n_{\psi'}| Q_{(n_1,n_2,n_3,n_4)} | n_{\psi'} \rangle
=\frac{n!}{n_1!n_2!n_3!n_4!} \\ \langle \psi'|Q_1|\psi'\rangle^{n_1}\langle \psi'|Q_2|\psi'\rangle^{n_2}
\langle \psi'|Q_3|\psi'\rangle^{n_3}\langle \psi'|Q_4|\psi'\rangle^{n_4}.
\end{multline}

The right-hand side of \eqref{distribution} coincides with the multinomial distribution, i.e. the  cumulative distribution of   $n$ {\em indipendent} random variables $\{\xi _j\}_{j=1,...,n}$ with $4$ attainable outcomes $h=1,2,3,4$ 
and elementary probabilities 

\begin{equation}
P(\xi_j=h)=\langle \psi'|Q_h|\psi'\rangle, \qquad j=1,...,n
\label{PhN-0}
\:.\end{equation}
The theoretical probabilities \eqref{PhN-0} are the building blocks for our further analysis.

\subsection{Estimate of the correlation coefficient}
The analysis presented so far shows that the photons in the beam can be considered as carriers of independent identically distributed random variables $\{\xi_j\}_{j\in \bN}$ with 4 possible outcomes and corresponding probabilities \eqref{PhN-0}, where
$$P(1)=  |\langle\psi'_{\textbf{a}, \textbf{b}}|1H \rangle|^2, \quad P(2)=  |\langle\psi'_{\textbf{a}, \textbf{b}}|1V \rangle|^2,\quad $$ $$ P(3)=  |\langle\psi'_{\textbf{a}, \textbf{b}}|OV \rangle|^2, \quad P(4)=  |\langle\psi'_{\textbf{a}, \textbf{b}}|OH \rangle|^2.$$ Denoting with $N^{(\textbf{a,b})}_{xy}$ (where $x=0,1$ and $y=V,H$) the numbers of counts on each measurement channel and $N^{(\textbf{a,b})}_{TOT}:=\sum_{x,y}N^{(\textbf{a,b})}_{xy}$, the relative frequencies 
\begin{equation}\label{frac} \frac{N^{(\textbf{a,b})}_{xy}}{N^{(\textbf{a,b})}_{TOT}} \end{equation}
provide an estimate of the elementary quantum-mechanical probabilities $|\langle\psi'_{\textbf{a}, \textbf{b}}|xy \rangle|^2$.

It is important to remark that the result is robust under non-idealities of the setup.
Indeed,  losses are not an issue provided that the four measurement channels, i.e. collecting optics and SPADs, have equal efficiencies.

Furthermore, by the independence  of the random variables $\xi_j$ (associated to the result of the measurement of single particle-observables on the photons composing the light beam)  there are no fundamental differences between an attenuated light and single (heralded) photons. In both cases,  we observe a flux of photons. In the case of the heralded photons, the arrival times can (to some extent) be determined by the experimental procedure, while, in the case of the attenuated light, the arrival times are stochastically distributed. In particular, under the assumption that the weights $C_n$ in the statistical mixture \eqref{state-i} are Poisson distributed, i.e.  $|C_n|^2=\frac{e^{-\mu}\mu ^n}{n!}$, the lapses of time  between two subsequent detections are independent and exponentially distributed with rate $\mu$. If the light intensity is sufficiently low to allow the SPAD to distinguish the single counts, then the correlation coefficient 
\begin{multline}\label{corr-coeff-0}
E(\textbf{a}, \textbf{b}) = |\langle\psi'_{\textbf{a}, \textbf{b}}|1V \rangle|^2 + |\langle\psi'_{\textbf{a}, \textbf{b}}|0H\rangle|^2 - \\ |\langle\psi'_{\textbf{a}, \textbf{b}}|0V \rangle|^2
	- |\langle\psi'_{\textbf{a}, \textbf{b}}|1H \rangle|^2,
\end{multline}
can be estimated as
\begin{equation}\label{corr-coeff}
E(\textbf{a}, \textbf{b}) = \frac{N^{(\textbf{a},\textbf{b})}_{1V} + N^{(\textbf{a},\textbf{b})}_{0H} - N^{(\textbf{a},\textbf{b})}_{0V} 
	- N^{(\textbf{a},\textbf{b})}_{1H}}{N^{(\textbf{a},\textbf{b})}_{0V} + N^{(\textbf{a},\textbf{b})}_{1H} + N^{(\textbf{a},\textbf{b})}_{1V} 
	+ N^{(\textbf{a},\textbf{b})}_{0H}}.
\end{equation}

\subsection{Meaning of Bell inequalities violation for weak and  intense light beams}\label{secintensebeam}

Let us focus attention to the $S$-parameter defined as 
\begin{equation}\label{eq:Sparam}
S(\textbf{a},\textbf{a'}, \textbf{b}, \textbf{b'}) := E(\textbf{a}, \textbf{b}) - E(\textbf{a}, \textbf{b'}) + E(\textbf{a'}, \textbf{b}) + E(\textbf{a'}, \textbf{b'}),
\end{equation}
where $E(\textbf{a}, \textbf{b})$ is the correlation coefficient \eqref{corr-coeff}.
Let us explain the relevance of this function in our experiment. The presence of {\em single photon detectors} in our experiment  allows to collect a time-ordered sequence $(x^{(\textbf{a})}_n,y^{(\textbf{b})}_n) $ (with $x^{(\textbf{a})}_n,y^{(\textbf{b})}_n\in \{+1, -1\}$) of outcomes of measurements of single particle observables $O^{\textbf{a}}=\textbf{a}\cdot \boldsymbol{\sigma}$ and $O^{\textbf{b}}=\textbf{b}\cdot \boldsymbol{\sigma}$, where the subscript $n$ stands for the chronological order of observation.  
If we describe the sequence of measurements outcomes $( x^{(\textbf{a})}_n ,y^{(\textbf{b})}_n)$  in terms of the realizations of  sequences of independent identically distributed discrete  random variables  $(\xi^{(\textbf{a})}_n, \xi^{(\textbf{b})}_n)$  and and we set   $E(\textbf{a,b})=\bE[\xi^{(\textbf{a})}\xi^{(\textbf{b})}]$ ($\bE[x]$ denotes the expectation value of the variable $x$), then a standard argument shows that the values of the $S$ parameter   satisfies   \begin{equation} |S(\textbf{a},\textbf{a'}, \textbf{b}, \textbf{b'})| \leq 2 \label{eq:BCHSH}\end{equation} for every choice of $\textbf{a}, \textbf{a'}, \textbf{b}$ and $\textbf{b'}$.
This is the celebrated
 CHSH inequality.
 Let us recast this result in physical terms making explicit some crucial physical hypotheses. 
 We assume that the beam is made of elementary constituents {\em already before} they are revealed by the single counts of the detectors. Evidence of single counts corroborates this hypothesis though it does not demonstrate it. We further suppose that these constituents are independent and that the measured quantities are properties of each of those constituents.
Within these overall assumptions,  if we also assume that (a) each outcome $( x^{(\textbf{a})}_n ,y^{(\textbf{b})}_n)$
corresponds  to pre-definite values  (realism) for the couple of  observables  $O^{\textbf{a}}$ and $O^{\textbf{b}}$ of the $n$-th revealed elementary constituent, and that (b) each value of
 $O^{\textbf{a}}$ is independent of the choice of ${\bf b}$ for the simultaneously measured observable  $O^{\textbf{b}}$ and {\em vice versa} (non-contextuality), then 
 the CHSH inequality \eqref{eq:BCHSH} must be valid. We stress that in this interpretation of the experimental data, the  values $x^{(\textbf{a})}_n$ and $y^{(\textbf{b})}_n$ are functions of some single-constituent  {\em hidden variable} $\lambda$ which randomly varies and it explains the observed stochasticity of the outcomes. In summary, under our hypothesis about the independent elementary constituents of the beam, failure of Bell inequality in the CHSH formulation \eqref{eq:BCHSH} rules out realistic non-contextual   hidden-variable explanations of the experimental data.
\begin{remark} {\em Without the said assumptions on the pre-existent independent constituents, one may try to construct a hidden variable theory where the beam is a classical wave, the clicks are nothing but an evidence of some quantum process in the detectors and the hidden variable says when a detector clicks.}\end{remark}
\begin{remark} {\em In addition, it is important to remark that the detectors we use have efficiencies of about 50 \%. Since we are actually sampling just half of the incoming photons, we have to adopt the fair sampling assumption, i.e. that the sample of detected photons is a representative of the emitted ones. By the way, this implies that our experimental results cannot rule out hidden variable theories where the choice of the detector to click or not is explicitly taken into account.}
 \end{remark}

Quantum mechanics produces a violation of inequality (\ref{eq:BCHSH}). The quantum mechanical corresponding of \eqref{corr-coeff-0} is
\begin{equation}
E(\textbf{a}, \textbf{b})  =  \cos(\phi-2\theta),
\end{equation}
where  $\phi$ and $\theta$ are the angles associated to the corresponding
 $\textbf{a}$ and $\textbf{b}$. Knowing the correlation coefficients, $S(\textbf{a},\textbf{a'}, \textbf{b}, \textbf{b'})$, i.e. $S(\phi,\phi',\theta,\theta')$, can be calculated.  
At a first glance, $S(\phi,\phi',\theta,\theta')$ seems a complex function of four parameters, but in fact only
three of the arguments are mutually independent. Indeed, the following equality holds:
\begin{equation}
\phi-2\theta =-\phi'-2\theta = \phi'-2\theta' = \alpha\:,\label{conditionalpha}
\end{equation} 
where $\alpha$ is a free parameter we can vary, giving
\begin{equation}
S(\alpha) = 3\cos \alpha - \cos(3\alpha)\:.\label{eq:SparamVartheta}
\end{equation}
The maximal violation of (\ref{eq:BCHSH}) 
foreseen by quantum mechanics  is  attained at $\alpha = \pm \pi/4$ where $S=  2\sqrt{2}$ and at $\alpha = \pm 3\pi/4$ where $S=  -2\sqrt{2}$ .

\begin{remark}
{\em The fact that the relative frequencies 
$N^{(\textbf{a,b})}_{xy}/N^{(\textbf{a,b})}_{TOT}$
in (\ref{frac})
give an estimate of the elementary quantum-mechanical probabilities \eqref{PhN-0}
 provides a purely quantum mechanical explanation of recent results \cite{Kagalwala2013,Aiello2015} where the violation of Bell inequalities is achieved even for intense light beams. According to our analysis, some phenomena considered as an example of the so-called  {\em classical entanglement}  can be actually traced back  to single-particle entanglement of the state of the single photons in the beam. Indeed,  by replacing the relative frequencies $N^{(\textbf{a,b})}_{xy}/N^{(\textbf{a,b})}_{TOT}$ with corresponding  light intensities $I^{(\textbf{a,b})}_{xy}/I^{(\textbf{a,b})}_{TOT}$, these still provide an estimate of the quantum-mechanical probabilities \eqref{PhN-0} and the correlation coefficients \eqref{corr-coeff} and \eqref{eq:BCHSH} will violate Bell inequalities for suitable choices of the parameters $\textbf{a, b , a',b'}$. 
 The crucial difference between our experimental context and the one described in \cite{Kagalwala2013} 
is the presence, in our case,   of single photon detectors. This allows to define a time-ordered sequence $(x^{(\textbf{a})}_n,y^{(\textbf{b})}_n) $ (with $x^{(\textbf{a})}_n,y^{(\textbf{b})}_n\in \{+1, -1\}$) of outcomes of measurements of single particle observables $O^{\textbf{a}}=\textbf{a}\cdot \boldsymbol{\sigma}$ and $O^{\textbf{b}}=\textbf{b}\cdot \boldsymbol{\sigma}$, instead of just the cumulative distribution $ N^{(\textbf{a,b})}_{xy}$ which, if alone, does not imply a single-constituent interpretation. Without the possibility of  single constituent interpretation the  violation of CHSH inequality is harmless.}
\end{remark}

\begin{remark}
{\em As  stressed in \cite{Markiewicz2019}, the possibility of violating Bell inequalities with multiple-particles states is a peculiarity of single-particle entanglement.  
Indeed, in the case of interparticle entanglement, i.e. between the same degree of freedom of  two different particles, the relative frequencies  $N^{(\textbf{a,b})}_{xy}/N^{(\textbf{a,b})}_{TOT}$ of the possible measurement outcomes   can be collected only when it is possible to clearly associate particles belonging to the same entangled pair. From an experimental point of view, this requirement can be fulfilled in terms of  a strict control of the arrival times. In the case where multiple-particle states are involved, it is only possible to collect the  marginal distributions, namely the numbers $N^{(\textbf{a})}_{x}=\sum _yN^{(\textbf{a,b})}_{xy} $ and $N^{(\textbf{b})}_{y}=\sum _xN^{(\textbf{a,b})}_{xy}$, which are not sufficient for the construction of the quantum-mechanical correlations \eqref{corr-coeff}. As mentioned in the introduction, the works  by Reid and Walls \cite{reid1986violations} and by {\.Z}ukowski et al. \cite{zukowski2016bell} are able to cope with these limitations proposing alternative Bell inequalities involving correlations between light intensities. }
\end{remark}

\section{The experiment}\label{Experiment}

\subsection{The experimental setup}\label{setup}

The setup of the experiment is schematically illustrated in fig. \ref{fig:setup}. The three light sources used for the measurements were:
\begin{itemize}
	\item An attenuated single mode green HeNe laser, emitting at $541$nm with nominal power of 5mW. The laser is fiber coupled and attenuated by a variable optical attenuator (VOA).
	\item  A commercial through-hole $5$mm LED with peak wavelength of $517$ nm and spectral width of $30$nm, which was filtered by an interference filter (IF) centered at $531$ nm with a bandwidth of $1$nm.
	\item  A Halogen lamp, model: HL-2000-FHSA-LL from Ocean Optics with a broad spectrum ($360-2400$nm), which was filtered at $531$nm to a $1$nm wide peak with the same interference filter used for the LED.
\end{itemize}
The sources were coupled to a single mode optical fiber. For the LED and the lamp, this ensures to collect photons from different spatial modes. A collimator at the end of the fiber feeds the input state to the generation stage. Here, a Glan-Thompson Polarizer (GTP) sets the light polarization to vertical. In this way, a photon is prepared in a definite momentum-polarization state. Then, a beam splitter (BS1) splits the signal in two different directions (momenta), i.e. the BS1 puts the photon in a superposition of momentum states. Half Wave Plates (HWP) are used to rotate the polarization by the indicated degrees. A piezoelectric transducer (PZT) actuated mirror controls a relative phase shift $\xi$ to compensate for any phase difference in the two arms. This operation is represented in fig. \ref{fig:intraparticle} by a wedge labeled PS, phase shifter. Thus, this first Mach-Zehnder Interferometer (MZI) performs a rotation of the input state in the momentum and polarization degree of freedom and generates the entangled state. Then, the light enters into a second MZI that transforms the entangled state to a specific state which is determined by the momentum phase $\phi$ and the polarization rotation angle $\theta$ of the two HWP. 

In the detection stage, two polarizing beam splitters (PBS) on each output arm of the second MZI separate the polarization and momentum states. The light is then coupled to long ($>1$ m) optical fibers and detected by four Si-SPADs (Excelitas). These long fibers are used to prevent cross-correlation, i.e. false counts, between the SPADs. The SPADs are here used since we do photon counting measurements. The counts from each SPAD are processed by a FPGA interfaced to a computer. The best detector efficiency was measured at $52\%$. We point out that this value does not allow a Bell test which closes the detection loophole, hence our hypothesis includes the fair sampling assumption.
Since the different detectors had different efficiencies we equalized their efficiencies by controlling the fiber-coupling of the signal to the detector. 
The typical dark count rates in our SPADs are few Hz, while the maximum signal measured at the exit of the generation stage is 2$\times$ 100 kHz. Since the characteristic dead time of our SPAD is 22 ns, this signal level is well within the dynamic range of the SPADs (linearity up to few MHz) and assure that we are dealing with weak signal, i.e. the probability that in a given time interval more than one photon reaches a detector is minimized.

The entire setup (sources and detector excluded) is enclosed in a dark box to isolate it from external environment and reduce the noise. The choice of the working wavelength is mainly motivated by the availability of both sources and detectors, but also by the possibility of having best performances (e.g. constant retardance of the polarizers and equal transmittance and reflectance of the beam splitters) for the optics. Since the signals are symmetric with respect to the chosen direction of measurement ($|0\rangle$ and $|1\rangle$ respectively), as described by the theory \cite{Gadway09}, the $S$-parameter was estimated by the projections over $|0 H\rangle$ and $|0V\rangle$ in the case of the LED and of the Halogen lamp. In the case of the laser, the $S$-parameter is estimated by acquiring all the four signals.

\subsection{The measurements}\label{measure}

\begin{figure}
	\includegraphics[width = 1 \linewidth]{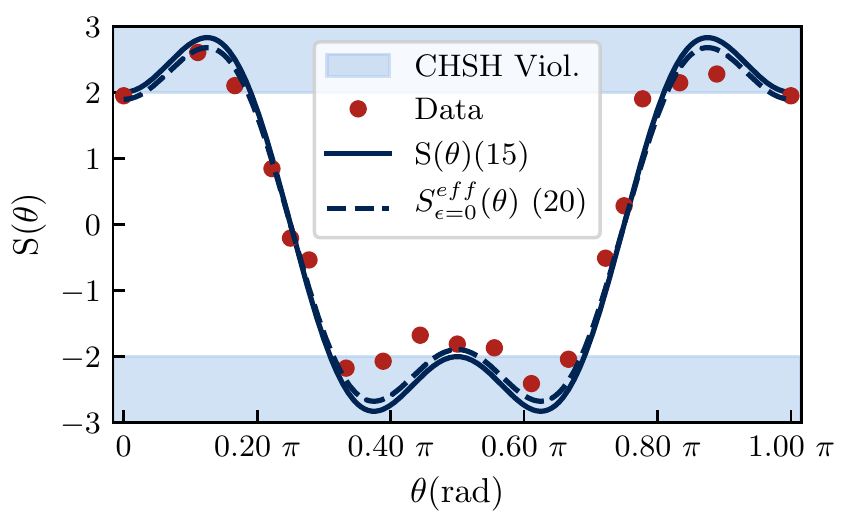}
	\caption{Violation of the CHSH inequality by an attenuated laser beam. Data (red points) are taken by varying the polarization angle $\theta$ of the HWPs in the preparation stage of the set-up of fig. 1. Error bars are within the dot size. The solid curve represents the $S$-parameter of eq. \eqref{eq:SparamVartheta}. The dashed curve is the $S^{eff}$-parameter given by \eqref{eq:Sepsilonvartheta} with $\eta=0.95\pm 0.01$ and $\epsilon=0$. The blue regions indicate the violation of the CHSH inequality \eqref{eq:BCHSH}.}
	\label{fig:CHSHlaser}
\end{figure}

The theoretical analysis of Sect \ref{secBell} is clearly reflected by our experimental results: by injecting light from the attenuated HeNe laser in the setup, we witness a violation of the Bell inequality, as can be seen in fig. \ref{fig:CHSHlaser}. For each data points, the average of several measurements is reported. Error bars are the errors propagated from the standard deviations of the measurements. If we fix $\phi=0$, we obtain $-2\theta=\alpha$ and the general expression of the $S$-parameter (\ref{eq:SparamVartheta}) takes the form  $S(\theta) = 3\cos (2\theta) - \cos(6\theta)$. Data are well reproduced by \eqref{eq:SparamVartheta}.

Since $E(\textbf{a}, \textbf{b})$ does not depend on whether the input state is a coherent superposition or a statistical mixture of pure states, SPE states can be generated also by attenuated incoherent sources, such as a lamp or a LED. In this case, the spontaneous nature of the photon emission results in a short coherence time $\tau_c$ and in a short coherence length $l_c$.
As discussed in Appendix, it is possible to  phenomenologically take into account the broad spectrum of any classical light source by replacing the entangled  state \eqref{eq:entangledstate} with the mixed state \cite{valles2014generation}
 \begin{equation}\label{eq:phenepsilon}
    \rho_\epsilon = (1-\epsilon) |\psi_{entangled} \rangle \langle \psi_{entangled}|+ \epsilon \rho_{Mixed}, 
\end{equation} 
where
\begin{eqnarray}
    \rho_{Mixed} &=& \frac{1}{2}|0H\rangle\langle 0H| + \frac{1}{2}|1V\rangle\langle 1V|,\label{state-i2}\\
      |\psi_{entangled}\rangle& =&\frac{1}{\sqrt 2}\left( |1V\rangle +  ie^{-iT\omega_0} |0H\rangle\right), \label{eq:es}
\end{eqnarray}
and $\epsilon$ is a phenomenological parameter which takes coherence properties into account. The pure state \eqref{eq:es} is a SPE that differs from \eqref{eq:entangledstate} by a relative phase which is a function of the time delay $T$ due to the different paths in the generation stage (fig. \ref{fig:setup}).
In the case of a source with a gaussian spectrum centered at $\omega_0$ and spectral half-width $\sigma_\omega$, quantum mechanics predicts that $\epsilon$ depends on $T$ as:
\begin{equation}\label{e(T)}
\epsilon(T) 
= 1 -e^{-T^2 \sigma^2_\omega/2}\:,    
\end{equation}
as explained in Appendix.
Since $\frac{1}{\sigma_\omega}=\tau_c$, \eqref{e(T)} shows that coherence is completely lost for $|T| \gg \tau_c$, where the state can be considered mixed and described by \eqref{state-i2}, while  for  $|T| \ll \tau_c$ the state \eqref{eq:phenepsilon} is entangled. From a practical perspective, the value of $\epsilon$ decreases by reducing $\sigma_\omega$ with spectral filtering and $T$ with optical alignment.

\begin{figure}
	\includegraphics[width = 1 \linewidth]{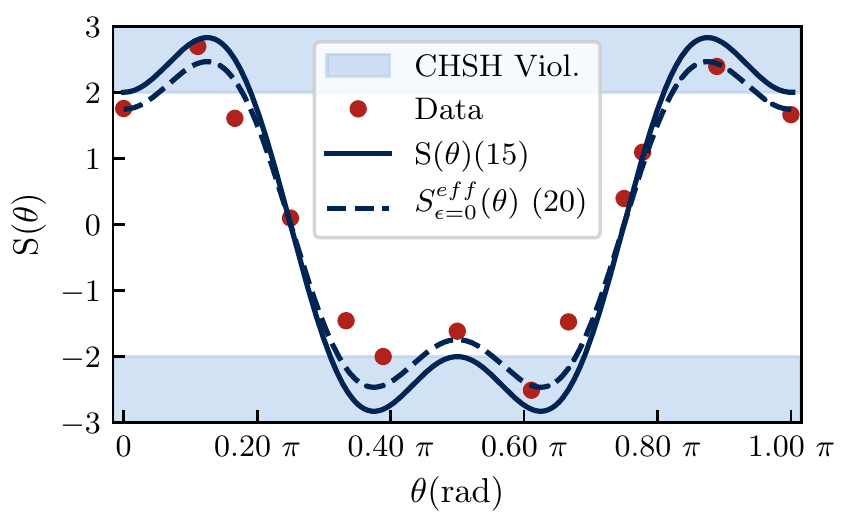}
	\caption{Bell inequality measurement for a filtered (1nm bandwidth) LED and for the setup aligned within the coherent regime. Data points are experimental results while lines are theoretical curves (full line ideal $S$-parameter \eqref{eq:SparamVartheta}, dashed line  $S^{eff}$-parameter \eqref{eq:Sepsilonvartheta} with $\epsilon=0$ and $\eta=0.87\pm0.02$). Error bars are within the dot size.}
	\label{fig:chshled}
\end{figure}

\begin{figure}
	\includegraphics[width = 1 \linewidth]{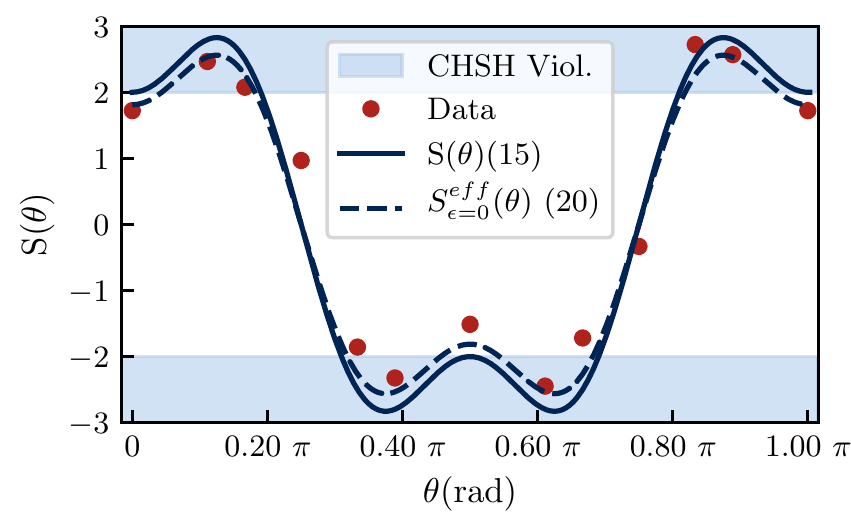}
	\caption{Bell inequality measurements for a 1 nm filtered halogen lamp. Data points are experimental results while lines are theoretical curves (full line ideal $S$-parameter \eqref{eq:SparamVartheta}, dashed line  $S^{eff}$-parameter \eqref{eq:Sepsilonvartheta} with $\epsilon=0$ and $\eta= 0.91 \pm 0.01$). Error bars are within the dot size.}
	\label{fig:chshthermal}
\end{figure}
\begin{figure}
	\includegraphics[width = 1 \linewidth]{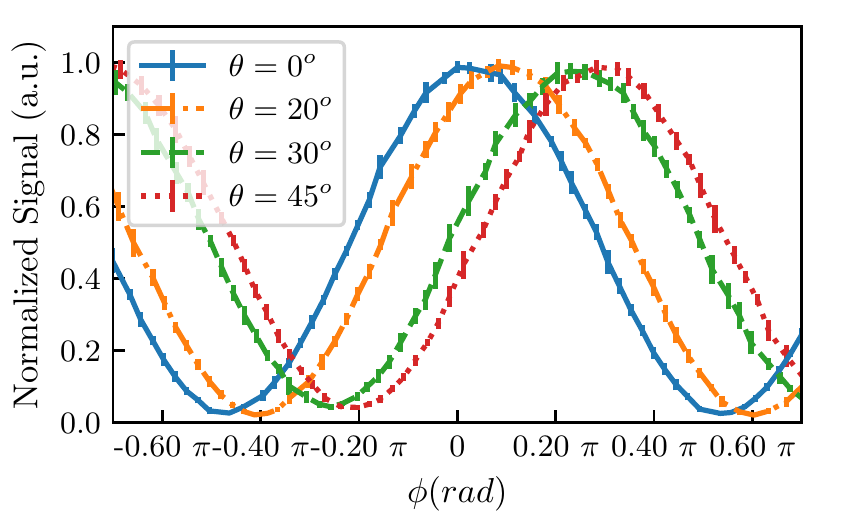}
	\caption{Signal acquired by one single SPAD ($N_{OH}^{(\phi,\theta)}$) as a function of $\phi$ for different polarization angles $\theta$ in the coherent case. Counts are normalized to the maximum count in the set.}
	\label{fig:visibility_coherent}
\end{figure}

In order to measure the single-particle entanglement generated by any classical light sources in the coherent regime, we performed the Bell inequality test by using two different filtered broadband sources: a halogen lamp and a LED. These are used as examples of light sources based on two different physical principles: black body radiation for the halogen lamp and spontaneous emission in semiconductors for the LED. Since the emission spectra of these sources are broad, we used a 1 nm interference filter to increase the source coherence length to $l_c$=$46\pm0.3 \mu$m. In figs. \ref{fig:chshled} and \ref{fig:chshthermal}, the $S$-parameter is reported as a function of the angle $\theta$. For specific values of $\theta$ we witnessed a violation of the CHSH inequality (\ref{eq:BCHSH}). As predicted by the theory,  the amplitude of $N^{(\textbf{a},\textbf{b})}_{0H}$ shows constant interference fringes for different values of $\theta$ (fig. \ref{fig:visibility_coherent}). Indeed, the visibility ($V=\frac{N_{OH}(max)-N_{OH}(min)}{N_{OH}(max)+N_{OH}(min)}$) does not change as a function of $\theta$.  In this case, \eqref{eq:SparamVartheta} can be generalized to include both incoherence by means of $\epsilon$ and noise, that reduces the visibility of the measurement channels, by means of a parameter $\eta $ (see Appendix):
\begin{multline}
S^{eff}(\theta)=\eta (1-\epsilon) \left(3\cos(2\theta)-\cos(6\theta)\right)+ \\ \eta \epsilon\left(  2\cos^3(2\theta)-2\sin^2(2\theta) \cos(6\theta)  \right)\:.
\label{eq:Sepsilonvartheta}\end{multline}

In fig. \ref{fig:chshled} and \ref{fig:chshthermal} we show that \eqref{eq:Sepsilonvartheta} reproduces the data with reasonable values of the noise and coherence parameters.\\

\begin{figure}
	\includegraphics[width = 1 \linewidth]{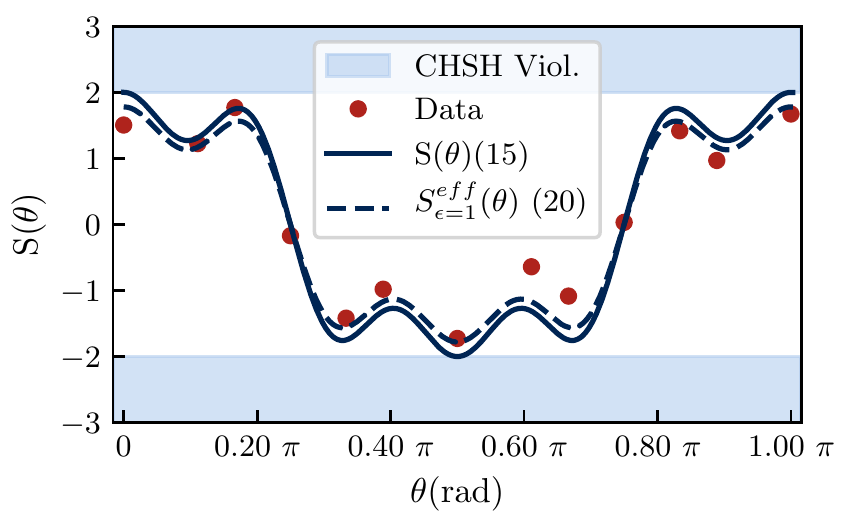}
	\caption{$S$-parameters (experimental data and theoretical curves) in the incoherent regime. $\eta=0.89\pm0.01$ and $\epsilon=1$ were used in the calculation of the  $S^{eff}$-parameter by \eqref{eq:Sepsilonvartheta}. Error bars are within the dot size.}
\label{fig:chshincoherent}
\end{figure}

\begin{figure}
	\includegraphics[width = 1 \linewidth]{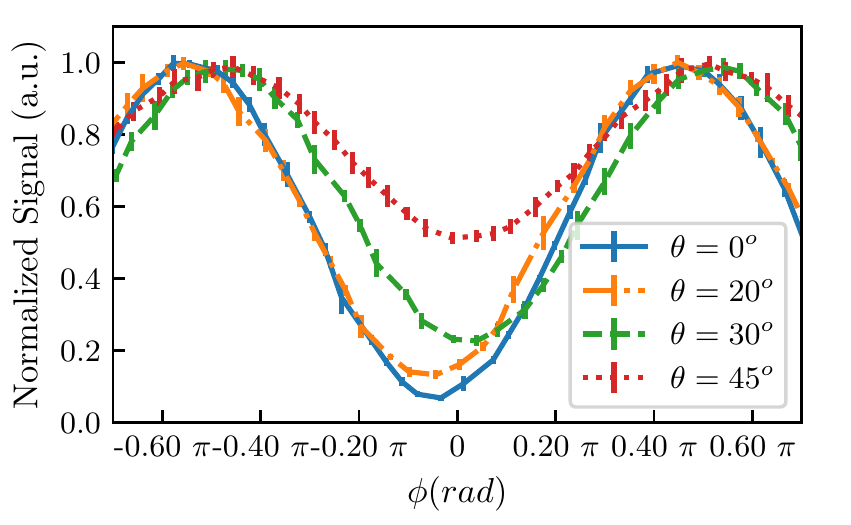}
	\caption{$N_{OH}^{(\phi,\theta)}$ as a function of $\phi$ for different $\theta$ for the incoherent case.}
	\label{fig:visibility_incoherent}
\end{figure}

To further confirm this theory, we repeated the measurements with the LED in the incoherent regime (figs. \ref{fig:chshincoherent} and \ref{fig:visibility_incoherent}). This is achieved by displacing the mirror in the first MZI of the set-up in fig. \ref{fig:setup} by a distance longer than $l_c$. As expected, the Bell inequality is never violated and the experimental points agree with the theoretical prediction for two independent states entering the measuring setup.  Here, the visibility in $N^{(\textbf{a},\textbf{b})}_{0H}$ changes as a function of $\theta$, approaching 0 for $\theta\approx \frac{\pi}{4}$. (fig. \ref{fig:visibility_incoherent}).

\section{Discussion}\label{discussion}

\subsection{Quantum signature}\label{QvsC}

As already remarked the peculiar correlations between different DoF of  a single photon has an  analogue in classical light beams.
This analogy, controversially called {\em classical entanglement}, was first raised  by  Spreeuw \cite{spreeuw1998} and concerns correlations between different DoF of a physical system whose nature is classical. The requirement is that it should admit a mathematical description in terms of tensor product spaces \cite{korolkova2019}.  A vector beam of light displaying  a non-uniform polarization 
pattern is a concrete elementary example of such a system as discussed in  \cite{berg2015,Aiello2015}. Here, the classical entanglement pops out between transverse spatial modes and polarization:  the electric field of a paraxial beam can be written as 
\begin{equation}\label{beam-cl-ent}
{\bf E}(\rho, z)={\bf e}_1f_1(\rho,z)+{\bf e}_2f_2(\rho,z)\:,
\end{equation}
where the unit vectors ${\bf e}_1,{\bf e}_2$  describe  the polarization and  the scalar functions $f_1,f_2$ describe the wavefront. 
 The propagation direction   is taken along $z$  whereas $\rho=x{\bf e}_1+y{\bf e}_2$ defines the transverse position vector. \\
The field \eqref{beam-cl-ent} is a vector in the tensor product $\cH_1\otimes \cH_2$, where  the generally complex vector spaces $\cH_1$ and  $\cH_2$ refer to the {\em classical} polarization and the {\em classical} spatial DoF, respectively. 
Via Schmidt decomposition \eqref{beam-cl-ent}  can be written as:
\begin{equation}\label{beam-cl-ent-2}
{\bf E}(\rho, z)=\sqrt{\lambda_1}{\bf u}_1g_1(\rho,z)+\sqrt{\lambda_2}{\bf u}_2g_2(\rho,z)\:.
\end{equation}
Above,  $\lambda_1,\lambda_2\in [0,1]$ and ${\bf u}_1,{\bf u}_2$ (resp. $g_1,g_2$) are orthonormal vectors in $\cH_1$ (resp. $\cH_2$), where 
\begin{equation}\label{orthogonal}\langle g_1,g_2\rangle _{\cH_2}=\int_{\bR^2}\bar g_1(\rho,z)g_1(\rho,z)d\rho=0.\end{equation}
If $\lambda_1\lambda _2=0$ then the field \eqref{beam-cl-ent-2} is {\em separable}, otherwhise it is called {\em classically entangled}. In particular, if $\lambda_1=\lambda_2=1/2$, then the field  \eqref{beam-cl-ent-2}
is mathematically equivalent to a maximally entangled Bell state of two qubits.  
The orthogonality condition \eqref{orthogonal} can be fulfilled if the functions $g_1,g_2$ have non-overlapping supports. In this case,  one obtains a {\em classical} version of the polarization-path entanglement.  Another possibility is to exploit first-order spatial modes of the electromagnetic field \cite{borges2010,gabriel2011}. In this case,
 \eqref{orthogonal} still holds even if the supports of $g_1$ and $g_2$ do actually overlap.

A  Bell test allows to certify the non-separability of the vector field \eqref{beam-cl-ent-2} as in the  quantum analogue. Given a couple of unit vectors $\ba, \bb$ associated to 
the joint measurement of the two DoF, the relevant correlation coefficient $E(\ba,\bb)$ is in this case
$$E(\ba,\bb)=  \frac{I_{++}^{(\ba,\bb)} + I_{--}^{(\ba,\bb)} - I_{+-}^{(\ba,\bb)} -I_{-+}^{(\ba,\bb)}}{I_{++}^{(\ba,\bb)} + I_{--}^{(\ba,\bb)} + I_{+-}^{(\ba,\bb)} + I_{-+}^{(\ba,\bb)}},$$
where  $I_{\pm\pm}^{(\ba,\bb)}$ are light intensites. Using $E(\ba,\bb)$ we can produce the $S$-parameter with the same structure of the quantum case (see Eq \eqref{eq:Sparam}).
A violation of the CHSH inequality is observed for suitable choices of the parameters $\ba,\bb$.
However, as the intensities are completely classical notions, this notion of ``classical entanglement'' and its properties can be completely discussed in the classical framework \cite{Khrennikov2020}. 
Even assuming the existence of quantum constituents of the classical wavepacket, the above  $S$-parameter cannot say anything about their nature. In fact, it  is a figure of merit which describes {\em collective} properties of the light beam, instead of the particular form of the state vector of the single photons.  
A complete classical description of the Bell inequality in this classical context is presented in \cite{Kagalwala2013}, where  the violation of the CHSH inequality is 
ascribed to particular coherence properties of the light beam.
We can assert that our theoretical analysis also provides a quantum mechanical interpretation of Bell inequalities violation for intense light beams in the case of non separability of spatial and polarization DoF. In these cases the discussion in Sect \ref{secintensebeam} and Appendix \ref{SEC1} provides
 an unified view of quantum SPE and “classical entanglement”: both can be ascribed to a particular form of the state of the single photons in the light beam (see \eqref{eq:phenepsilon}). Even if $E(\textbf{a}, \textbf{b})$ and the $S$-parameter are evaluated in terms of classical quantities (e.g. as intensities of light in \cite{Kagalwala2013}), they still admit a quantum mechanical interpretation. In this picture, the coherence properties of the light can still be taken into account by means of \eqref{eq:phenepsilon} and \eqref{e(T)}, so that a violation of the Bell inequality is related to first-order coherence and can be interpreted as single-particle entanglement of the one-photon states.
 
The point of view of this paper is however that he violation of the Bell inequality must be  ascribed to properties of  the single photons when the experiments are able to highlight the particle nature of the light, i.e., when experimental data regard {\em counts of photons rather than  intensities}. Indeed, measurements are now made on {\em each  single constituent} of the beam(s) and the violation of the CHSH inequality rules out the interpretation of the statistical results in terms of (classical) realistic non contextual theories of each single constituent. The situation has some similarities with  the phenomenology of  Young's experiment. As soon as we are no longer able to discriminate the  spots on the screen due to single photons, everything can be described with the classical theory of light: the physical object is a (classical) beam and we are measuring its intensity. If  we are instead able to 
observe the single spots, realizing that the interference figure is nothing but a superposition of an enormous number of them,  we are forced to interpret all the observed phenomenology in terms of these elementary constituents. The properties of these elementary constituents cannot be explained in terms of classical electromagnetism.

Within the interpretation of non-locality as a basic property of quantum entanglement, a possible argument against the quantumness of SPE is that the violation of Bell inequalities, in absence of spatial separation, cannot be interpreted as a signature of non-locality and consequently as a signature of the quantumness of the correlations. However this viewpoint does not definitely undermine the quantum nature of SPE because the violation of Bell inequalities in this case can be interpreted as a signature of \emph{quantum contextuality} \cite{Markiewicz2019}. In addition, when discrete events (clicks of detectors) come into play, the Bell inequalities violation provides a quantum signature and can be ascribed to non-commutativity of couples of observables \cite{Khrennikov2020}.

\subsection{Technological impact}

From an experimental point of view, the use of SPE is advantageous with respect to two-particles entanglement. In fact, to observe a violation of Bell inequalities with two-particles entanglement the knowledge of the joint distribution of the outcomes  $N^{(\textbf{a,b})}_{xy}$ is needed. These data are available only if we are able to recognize when  a pair of  measurements is referred to a single entangled pair. In other words, a strict control of the pair arrival times is fundamental. This experimentally demanding requirement is not necessary in the case of SPE because, in this case, the numbers $N^{(\textbf{a,b})}_{xy}$ can be simply obtained by looking at the number of counts in the four different channels of the experimental setup. That is why the times of arrival do not play a fundamental role and Bell inequalities are violated even with intense light beams (see Sect \ref{secintensebeam}).
Our results confine the need of expensive sources of single photons, e.g. heralded photons, to those Quantum Information protocols that require a deterministic time of arrivals of the single photons.
Even if the statistics of the input light does not affect the SPE, it must be taken into account in  practical applications to Quantum Information tasks. For example, SPE has been suggested for increasing the security of BB84-type QKD protocols \cite{Adhikari15}. In fact, security check based on violation of Bell-type  inequalities increases the robustness against side-channel attacks. On the other hand, one should also consider that thermal sources feature a super-poissonian emission statistics, and the presence of multiphoton components opens the risk of photon number splitting attacks. 

Other applications of SPE can be in the implementation and certification of quantum random number generators (QRNG). Indeed, the violation of Bell inequalities allows to prove a lower bound for the entropy of the random sequence produced \cite{pironio2010random}. In this case, the photon statistics plays no role. The presence of multiphoton states does not affect the randomness of the results as the system acts independently on each photon. The only consequence is the possibility of coincidence counts in the detectors, which cannot contribute to the random string and, therefore, decreases the generation rate. However, by using classical attenuated sources, the number of coincidence counts over the total signal can be easily controlled by the source intensity. Though, the signal-to-coincidence ratio is expected to be worse for incoherent light sources than for an attenuated laser. Therefore, using incoherent light compromises the performances with the feasibility (cost, size, power consumption, weight) of the QRNG compared to  a laser, but it may still be a good trade-off given the advantage of using an incoherent light source.\\

\section{Conclusions}\label{conclusions}
From an experimental point of view, we  have demonstrated that SPE can be generated from attenuated coherent and classical light sources and that SPE from such sources does indeed violate the Bell inequality. The crucial condition for SPE to be observed is that self-coherence (first-order coherence) between the involved DoF is preserved. We have also argued that single-particle or two-particles entanglement violations of the Bell inequalities are interpreted as a signature of a non-classical nature of the measured system: contextuality or non-locality, respectively \cite{Markiewicz2019}. Finally, we have shown that quantum states of light can be generated from cheap, compact, and low power photon sources. Contrary to common believe, quantum optics can be performed with a simple LED and does not need expensive and high power lasers.

\begin{acknowledgements}
We acknowledge helpful discussions with P. Bettotti in the initial phase of the experiment and with S. Azzini on single-particle entanglement. G. Fontana developed the acquisition system based on FPGA. We also thank M. Zukowski for having pointed out to us ref. \cite{zukowski2016bell}. This project has received funding from the European Union’s Horizon 2020 research and innovation programme under grant agreement No 820405 project QRANGE, and by the India-Trento Programme of Advanced Research ITPAR phase IV project. The work of N.L. was supported by a Q@TN grant and the one of D.P. by Fondazione Caritro. 
\end{acknowledgements}

\appendix*

\section{Analysis of the coherence length/time}\label{SEC1}
\begin{figure}
	\centering
	\includegraphics[width = \linewidth]{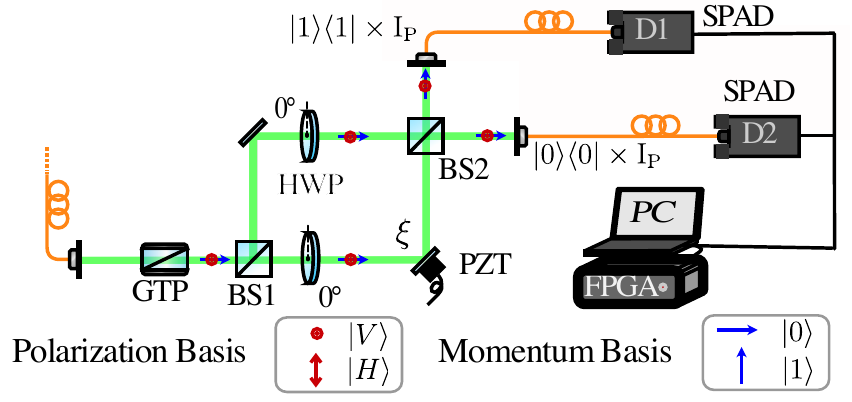}
	\caption{Setup for the autocorrelation measurements}
	\label{fig:Presetup}
\end{figure}
\begin{figure}
		\includegraphics[width = 1 \linewidth]{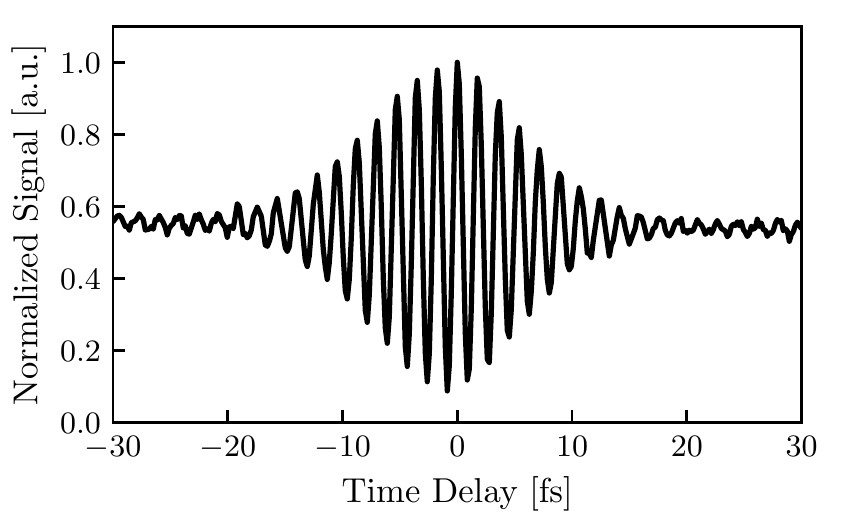}
	\caption{Autocorrelation of the unfiltered LED. It is acquired by moving the piezoelectric transducer by $20\mu$m.}
		\label{fig:AutoNF}
	\end{figure}
\begin{figure}
		\includegraphics[width = 1 \linewidth]{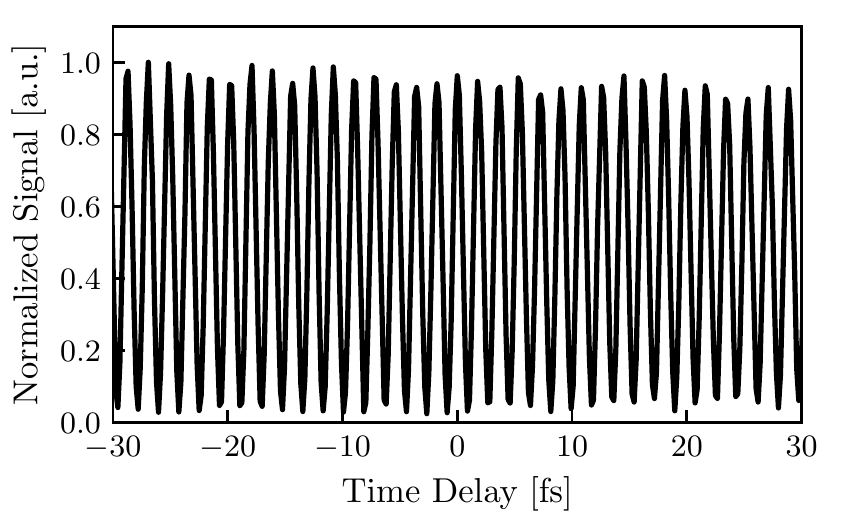}
		\caption{Autocorrelation of the filtered LED.}	\label{fig:AutoF}
	\end{figure}

Before the initial generally  multi-particle state (either $|\Psi\rangle $ or $\rho$ as in Sect. \ref{sec1})  encounters the final stage of detection, 
it is transformed by the intermediate stage of the circuit. As all elements of the circuit are {\em linear},
the net action on the multi-particle 
state directly arises from the action on a single-particle state as is discussed in Sect. \ref{Secmanyparticles}. We stick to 
the analysis of the action of a one-particle state and to some issues concerning temporal and spatial coherence which can be tackled at one-particle level. 
Since the source is not perfectly monochromatic and has a finite coherence length/time,  the finite-dimensional description 
of the one-particle photon states as presented in Sect. \ref{secBell} by means of the space of the states $\mathcal{H}_S = \mathcal{H}_M \otimes \mathcal{H}_P$
is correct as long 
as the  difference between the  length of the  arms of  the  MZI is less than the coherence length of the light source or, equivalently, the accumulated delay between the two photons belongs to the interval of coherence time.  
In a preliminary test we measured the autocorrelation of the LED source to analyze its coherence properties. To do so, we used the setup represented in Fig. \ref{fig:Presetup}.  The input state is injected to the generation stage by the use of an optical fiber and a collimator. Here, a Glan-Thompson Polarizer (GTP) sets the light polarization to vertical. Then a beam splitter (BS1) splits the signal in two different directions (momenta). A piezoelectric transducer (PZT) controls the relative phase shift $\xi$ between the two arms. At the output of the second beam splitter (BS2), the state is superposed over the two possible momentum states, signals of which, are acquired by the use of two Single Photon Avalanche Diodes (SPADs). These are connected to an FPGA interfaced to a computer.
 The emission spectrum of the LED can be approximated by a gaussian, and we fitted it with the function $ f(\omega) = Ae^{-\frac{(\omega-\omega_0)^2}{2\sigma^2_\omega}}$.
From this fit we obtain $A=0.932 \pm 0.002$, $\omega_0 = (3611.4 \pm 0.4)$THz and $\sigma_\omega = (134\pm 9)$THz. This gives $\tau_c =\frac{1}{\sigma_\omega} =(7.43 \pm 0.02)$fs, or a coherence length of $l_c =\tau_c c =(2.227 \pm 0.006) \mu$m. In order to increase the coherence length of the LED, we filtered it by a $1$nm interference filter centered at $531$nm. The filtered spectrum is also fit with a gaussian with parameters $A=(0.985 \pm 0.006)$, $\omega_0=(3547.24 \pm 0.04)$THz and $\sigma_\omega = (6.5\pm 0.8)$THz. In this way we obtain $\tau_c = (154 \pm 1)$ fs and $l_c = (46.0 \pm 0.3) \mu$m. The difference coming from the increase in the coherence time can be seen by comparing the filtered and unfiltered autocorrelations (see figs. \ref{fig:AutoNF} and \ref{fig:AutoF}). Here we report the autocorrelation as a function of the time delay between the two optical paths. Note that the oscillations of the signal decrease rapidly for the unfiltered LED while they do not decrease significantly over the $20\mu$m range for the filtered LED.

\subsection{A  precise description of the one-particle state and coherence length/time}
The state of a single  photon exiting from the first beam splitter (Fig. (\ref{fig:Presetup})) is actually a (superposition) of normalized packets $\psi_j({\bf k}) \otimes |\theta\rangle$, below simply denoted by  $\psi_j({\bf k}) |\theta\rangle$,
where 
\beq\label{theta} |\theta\rangle := \cos \theta |V\rangle + \sin \theta |H\rangle\eeq  is the polarization part of the state and the function $\psi_j$ is sharply concentrated around the
value ${\bf k}_j \in \mathbb{R}^3$.
As we know, these momenta $\hbar {\bf k}_0,  \hbar {\bf k}_1$ define the sharp states $|0\rangle, |1\rangle$ which are a rough but effective
approximation of the 
functions $\psi_0$ and $\psi_1$ we exploited to describe the space of the states as in Sect.\ref{sec1}.
It holds $\omega_0 = c|{\bf k}_0|=c|{\bf k}_1|$ where $\frac{\omega_0}{2\pi}$ is the frequency of the filtered light entering 
the whole setup. The functions $\psi_j$, with $j=1,2$, vanish outside two corresponding small balls, respectively,  $B_j \subset \mathbb{R}^3$,  
whose centers are the vectors ${\bf k}_j$
and such that $B_1$ and $B_2$ are sharply disjoint. 
The appearance of a coherence length (or coherence time) is here explained in terms of 
time evolution of the wave functions
\beq U_t \psi({\bf k}) |\theta \rangle = e^{-i c|{\bf k}| t} \psi({\bf k})  |\theta\rangle\:. \label{UE}\eeq
It is clear that, within this more precise view, the Hilbert space can no longer be considered finite dimensional since we have  different states  at different times because of the dependence of  the phase  $e^{-i c|{\bf k}| t}$ 
 on ${\bf k}$. 

The entangled state state entering  the second beam splitter in Fig. (\ref{fig:Presetup}) is no longer the Bell state
\begin{equation}\frac{1}{\sqrt 2}\left( |1 V\rangle +  i|0H\rangle \right),\label{state-ideal}\end{equation} 
but:
\beq  |\psi^{\scriptsize\mbox{(precise)}}_{\scriptsize\mbox{entangled}} \rangle :=\frac{1}{\sqrt{2}} 
\left( \psi_1({\bf k})|V\rangle + i e^{-icT|{\bf k}|} \psi_0({\bf k})|\theta \rangle\right). \label{prent} \eeq
An overall  phase $e^{-ict|{\bf k}|}$ irrelevant for our computation will henceforth be omitted. 
Here  $T$ is the delay time between
the two paths exiting from the first beam splitter, i.e. $cT|{\bf k}|=\phi$ is therefore the phase difference acquired during the propagation in the two different arms.  In order to quantify  the robustness of the single-particle entanglement, we measure the interference between the two addends of the 
state through the final stage of the preliminary experiment consisting of a second beam splitter and a pair of detectors   as in Fig. (\ref{fig:Presetup}).

The state entering the detectors exiting the second beam splitter is described as follows,
taking the unitary transformation describing the second beam splitter into account. 
\begin{multline}
|\psi^{\scriptsize\mbox{(precise)}}_{out} \rangle =(U_{BS}\otimes I)|\psi^{\scriptsize\mbox{(precise)}}_{\scriptsize\mbox{entangled}} \rangle\\
= \frac{1}{2}(i\psi_0({\bf k})+\psi_1({\bf k})) |V\rangle+\frac{i}{2}e^{-ict|\bf k|}(\psi_0({\bf k})+i\psi_1({\bf k})) |\theta\rangle\label{cs} 
\end{multline}

With the sharp packet approximations, the detectors D1 and D2  at the end of  the circuit represented in Fig. (\ref{fig:Presetup}) are mathematically described by orthogonal projectors $|0\rangle \langle 0|$ and  
$|1\rangle \langle 1|$. 
However, moving on to the more accurate description, we can represent those orthogonal projectors as the multiplicative operators 
$P_j := \chi_{K_j}({\bf k})$ in the space of momentum packets,
where $\chi_{K_j}({\bf k})=0$ if ${\bf k} \not \in K_j$ and $\chi_{K_j}(\bf k) =1$ if ${\bf k}\in K_j$.  Here $K_j$ is 
a set of momenta which necessarily  includes  the corresponding  
ball $B_j$ and such that
$K_0\cap K_1 = \emptyset$. While the extension 
of $B_j$ is decided by the source, the shape of $K_j$ is  fixed by the detector. An expected  shape of $K_j$ is a truncated
cone whose axis is parallel to ${\bf k}_j$ and whose bases are 
portions of parallel spherical surfaces whose distance from the origin of the space of momenta is respectively proportional to
the minimal and the maximal frequency detectable by the device.
The probability to detect the photon in the $j$-th detector with a polarization selected by the one-dimensional
orthogonal projector $Q$ acting in $\bC^2_{\scriptsize\mbox{polarization}}$ is
$\langle\psi^{\scriptsize\mbox{(precise)}}_{out} |P_j \otimes Q| \psi^{\scriptsize\mbox{(precise)}}_{out}\rangle\:.$
We have for $j=0,1$,
\begin{multline}
\langle\psi^{\scriptsize\mbox{(precise)}}_{out} |P_j \otimes Q| \psi^{\scriptsize\mbox{(precise)}}_{out} \rangle= \\
\label{nuovo} \frac{1}{4}\Big[\langle V|Q|V\rangle+\langle \theta| Q|\theta\rangle+ \\(-1)^j 2  Re\Big(\langle V|Q|\theta\rangle \int_{\bR^3}|\psi_j({\bf k})|^2 e^{-icT|{\bf k}|} d^3k\Big)\Big]\:.
\end{multline}

In particular, for $Q=I$:

\begin{multline}
\label{precise}\langle \psi^{\scriptsize\mbox{(precise)}}_{out} |P_j \otimes I| \psi^{\scriptsize\mbox{(precise)}}_{out} \rangle= \\
  \frac{1}{2} \left(1 + (-1)^j\cos \theta   \int_{\bR^3} \cos(cT|{\bf k}|) |\psi_j({\bf k})|^2 d^3k\right)\:.
\end{multline}
When dropping the third addend in the parenthesis of (\ref{nuovo}) and the second in (\ref{precise}) we obtain the same respective  results as  that of an incoherent superposition
\beq \rho_{Mixed} = \frac{1}{2} \left( |1\rangle \langle 1|\otimes  |V\rangle\langle V| + 
|0\rangle \langle 0|\otimes  |\theta\rangle\langle \theta|\right)\label{is}\eeq
entering the second detector  in place of the entangled state $ |\psi^{\scriptsize\mbox{(precise)}}_{\scriptsize\mbox{entangled}}\rangle$ defined in 
(\ref{prent}).
The second  addend in the right-hand side of (\ref{precise}) and the third in the right-hand side of (\ref{nuovo})  describe the quantum interference as a function of the  delay time $T$. Let us focus on the simplest case of $Q=I$ in (\ref{precise}),  since the general  case is a trivial extension of this.  It holds
$$ \int_{\bR^3} \cos(cT|k|) |\psi_j({\bf k})|^2 d^3k  = Re \int_0^{+\infty} e^{iT\omega} f(\omega) d\omega\:, $$
where   $Re \, z$ denotes the real part of  $z\in \bC$ and
\beq\label{precise2}f(\omega) =\frac{\omega^2}{c^3}\int |\psi_j( \omega/c, \vartheta,\varphi )|^2  \sin \vartheta d\vartheta d\varphi. \eeq 
The right-hand side is nothing but $|\psi_j({\bf k})|^2$ integrated only along the two polar angles  
$\vartheta, \varphi$ of the vector ${\bf k}$ 
whose norm is $\omega/c$. We are assuming here  that the
function $f$ does not depend on $j$, since we are approximating it with a Gaussian sharply concentrated on $\omega_0= c|{\bf k}_0|= 
c|{\bf k}_1|$
with a standard deviation $\sigma_\omega$ which is nothing but  the width of the filter for the entering light,
\beq \label{precise3}
f(\omega) =  \frac{e^{-\frac{(\omega -\omega_0)^2}{2 \sigma^2_\omega}}}{ \sqrt{2\pi \sigma^2_\omega}} \:.
\eeq
Since $\omega _0\gg\sigma_\omega$, we can estimate the integral as
$$ Re \int_0^{+\infty} e^{iT\omega} f(\omega) d\omega \simeq Re \int_{-\infty}^{+\infty} e^{iT\omega} f(\omega) d\omega = $$ $$ Re (e^{iT\omega_0} g(T))=\cos(\omega_0T) g(T)$$
where $g(T)=e^{-\frac{1}{2}\sigma^2_\omega T^2}$. 
The shape of this function is qualitatively identical to fig. \ref{fig:AutoNF}. The function $g$ is, up to normalization terms,  the density of a centered Gaussian measure with 
 standard deviation given by $\sigma_T =1/\sigma_\omega$.
Therefore, if $|T| \gg 1/ \sigma_\omega$,  the interference term is negligible in (\ref{precise}) and the state (\ref{cs}) can be safely 
replaced by the incoherent superposition (\ref{is}) as we shall better discuss in the next section.
When the delay $T$ is inside the coherence region, it makes sense to come back to the initial rougher approximation
(see the next section)
of very sharply peaked packets, approximating the phase $e^{-icT|k|}$ in (\ref{precise}) with $ e^{-iT\omega_0}$, i.e.  
$\xi = -T\omega_0$,
and defining  the entangled state 
entering the second beam splitter as
\beq  |\psi_{\scriptsize\mbox{entangled}} \rangle :=\frac{1}{\sqrt{2}} 
\left( |1V\rangle + i e^{i\xi} |0 \theta \rangle\right) \label{prent2} \eeq
so that, the state exiting from the second beam splitter is
$$ |\psi_{out} \rangle = \frac{i}{2}|0\rangle  \left[ (1+ e^{i\xi}\cos \theta) |V\rangle +e^{i\xi} \sin \theta |H\rangle \right]$$
\beq + \frac{1}{2} |1\rangle  \left[ (1- e^{i\xi}\cos \theta) |V\rangle - e^{i\xi} \sin \theta |H\rangle \right]\:.\label{cs2} \eeq
finding

$$ \langle \psi_{out}|P_j \otimes I| \psi_{out}\rangle=  \frac{1}{4} \left[|1+ (-1)^je^{i\xi}\cos \theta|^2 + 
\sin^2\theta\right] = $$ \beq \frac{1}{2}(1+ (-1)^j\cos \theta \cos \xi) \:.\label{P22}\eeq

\subsection{Effective analysis of the coherence length/time}\label{seccoherenceeff} To complete the discussion about the length/time coherence,  
we show how our picture  is in agreement -- and actually explains it --  with the phenomenological
description proposed by \cite{valles2014generation} where the action of the 
state 
$|\psi^{\scriptsize\mbox{(precise)}}_{entangled} \rangle \langle \psi^{\scriptsize\mbox{(precise)}}_{entangled} |$ 
when  measuring relevant observables 
is effectively 
represented in terms of  a  phenomenological density matrix $\rho_\epsilon$ in the $4$-dimensional Hilbert space 
$\bC^2_{\scriptsize momentum} \otimes \bC^2_{\scriptsize polarization}$
as follows
\beq\label{rhoep}
\rho_\epsilon = (1-\epsilon) |\psi_{entangled} \rangle \langle \psi_{entangled}|+ \epsilon \rho_{Mixed}
\eeq
where $\rho_{Mixed}$ is given by (\ref{is}), $|\psi _{entangled}\rangle$ is the pure vector in $\bC^2\otimes \bC^2$ defined in (\ref{prent2})
where we make explicit here  the relation $\phi= -T\omega_0$ for future convenience,
\begin{equation}|\psi_{entangled}\rangle =\frac{1}{\sqrt 2}\left( |1 V\rangle +  ie^{-iT\omega_0} |0\, \theta\rangle \right) \:.
\label{state-ideal-2}\end{equation}
Finally, $0 \leq \epsilon \leq 1$ is a phenomenological parameter which takes coherence properties into account: $\epsilon=0$ means that we are inside the coherence length, while $\epsilon=1$ means that coherence is lost. 

 We expect that $\epsilon$ is a function of the delay time $T$ and we go to investigate this relation with the help of a more precise analysis of the coherence than the one performed in the previous section.

The effective state of the system, after the action of the second beam splitter, is given by 
$$\rho_{\epsilon,out}=(U_{BS}\otimes I)\rho_\epsilon(U_{BS}\otimes I)^\dagger.$$

What really matters in our approach are the expectation values  of observables
$|j\rangle\langle j| \otimes Q_j$ in the Hilbert space $\bC^2_{\scriptsize momentum} \otimes \bC^2_{\scriptsize polarization}$ corresponding 
to observables $P_j \otimes Q$ in the infinite dimensional Hilbert space where the precise time evolution (\ref{UE}) takes place,  where $Q$ works
in the polarization space. Hence, $\rho_{\epsilon,out}$ is supposed to satisfy
$$tr(\rho_{\epsilon,out} |j\rangle\langle j| \otimes Q)=  \langle\psi^{\scriptsize\mbox{(precise)}}_{out} |P_j \otimes Q| 
\psi^{\scriptsize\mbox{(precise)}}_{out} \rangle$$
and identity (\ref{rhoep}) has to be interpreted as 
\begin{multline}
\langle\psi^{\scriptsize\mbox{(precise)}}_{out} |P_j \otimes Q| \psi^{\scriptsize\mbox{(precise)}}_{out} \rangle =\\ = (1-\epsilon)
\langle \psi_{entangled} | (U_{BS}\otimes I)^\dagger(|j\rangle \langle j| \otimes Q) (U_{BS}\otimes I)|\psi_{entangled}  \rangle \\ +\epsilon  \:\mbox{tr}[(U_{BS}\otimes I)\rho_{Mixed}(U_{BS}\otimes I)^\dagger(| j\rangle \langle j|\otimes Q)]
\end{multline}
Expanding the left-hand side of the identity above, with $\langle V| Q|\theta\rangle =|\langle V|Q|\theta\rangle| e^{i\varphi}$, 
taking (\ref{precise}) into account and exploiting  (\ref{is}) and (\ref{P22}) in the right-hand side, taking the made approximations into account and  (\ref{precise2}), (\ref{precise3}) in particular,   we eventually  find the relation linking $\epsilon$ to $T$ and $\sigma _\omega$,
\begin{equation}\label{epsilonT}\epsilon(T) 
=  1 - \int_{-\infty}^{+\infty} \frac{\cos(\omega T+\varphi) }{\cos(\omega_0T+\varphi)}  \frac{e^{-\frac{(\omega -\omega_0)^2}{2 \sigma^2_\omega}}}{ \sqrt{2\pi \sigma^2_\omega}}  d\omega = 1 -e^{-T^2 \sigma^2_\omega/2}\:.\end{equation}
As expected,  coherence is completely lost for 
$|T| \gg \frac{\sqrt{2}}{\sigma_\omega}$, there  the state can be considered mixed and described by (\ref{is}) in that regime.
For  $|T| \ll \frac{\sqrt{2}}{\sigma_\omega}$  the approximated state (\ref{cs2}) can be safely used  in place of the more accurate one (\ref{cs}).

\subsection{CHSH inequality for partially incoherent states}\label{secAdd} 
We want to compute here the value of the CHSH parameter $S$ when the measured state is the partially incoherent one arising from 
(\ref{rhoep}) with the value $\epsilon$ compatible with the experimental setup. The discussion of Section \ref{secBell} can be restated 
simply replacing the pure  state $|\Psi_+\rangle$ entering the stage (II) of the setup with the mixed state 
\begin{equation}\rho_\epsilon = (1-\epsilon) |\Psi_+\rangle \langle \Psi_+| + \epsilon \rho_{Mixed}\label{MRHO}\end{equation}
with $\rho_{Mixed}=\frac{1}{ 2}(|0H\rangle\langle 0H|+|1V\rangle\langle 1V|)$.
The $\phi$ phase shift in the second MZ induces a rotation $U(\phi/2)$ of an angle $\phi/2$ in the momentum Hilbert space. Setting $\tilde \phi=\phi/2$ we have: 
\begin{multline}
U(\tilde\phi)|0\rangle=\cos\tilde\phi |0\rangle- \sin\tilde\phi |1\rangle, \\
\qquad U(\tilde\phi)|1\rangle=\cos\tilde\phi |1\rangle+ \sin\tilde\phi |0\rangle.\label{UUU}\end{multline}
The polarization rotators in the final stage of the setup produce a rotation $U(\theta)$  in the polarization Hilbert space of an angle $\theta$:
\begin{multline} U(\theta)|V\rangle=\cos\theta|V\rangle+\sin\theta |H\rangle, \\ \qquad U(\theta)|H\rangle=\cos\theta |H\rangle-\sin\theta |V\rangle.
\label{UUUU}\end{multline}
By setting
$\rho'=U(\tilde\phi)\otimes U(\theta)\rho_{Mixed} (U(\tilde\phi)\otimes U(\theta))^\dag$ we obtain the following probabilities
\begin{eqnarray*}
	P_{0V}&=&Tr[\rho'|0V\rangle\langle 0V|]=\frac{1}{2}(\cos^2\tilde\phi\sin^2\theta+\sin^2\tilde\phi\cos^2\theta)\\
	P_{0H}&=&Tr[\rho'|0H\rangle\langle 0H|]=\frac{1}{2}(\cos^2\tilde\phi\cos^2\theta+\sin^2\tilde\phi\sin^2\theta)\\
	P_{1V}&=&Tr[\rho'|1V\rangle\langle 1V|]=\frac{1}{2}(\cos^2\tilde\phi\cos^2\theta+\sin^2\tilde\phi\sin^2\theta)\\
	P_{1H}&=&Tr[\rho'|1H\rangle\langle 1H|]=\frac{1}{2}(\cos^2\tilde\phi\sin^2\theta+\sin^2\tilde\phi\cos^2\theta)
\end{eqnarray*}
and 
$$P(\tilde\phi, \theta)=P_{1V}+P_{0H}-P_{0V}-P_{1H}=\cos(2\tilde\phi)\cos(2\theta).$$
Hence, we obtain
\begin{multline*}
	S(\tilde\phi,\tilde\phi',\theta, \theta') = \\P(\tilde\phi, \theta)+P(\tilde\phi', \theta)+P(\tilde\phi', \theta')-P(\tilde\phi, \theta')\\
	=\cos(2\tilde\phi)\cos(2\theta)+\cos(2\tilde\phi')\cos(2\theta)+ \\ \cos(2\tilde\phi')\cos(2\theta')-\cos(2\tilde\phi)\cos(2\theta').
\end{multline*}
In the case where $\tilde\phi=0$, $\theta=\alpha$, $2\tilde\phi'=2\alpha$, $2\theta'=3\alpha$, the corresponding 
value of $S$ is
$$ S^{Mixed}=2\cos^3\alpha-2\sin^2\alpha \cos(3\alpha) \:. $$
Dealing with the entangled state $|\Psi_+\rangle$ only as in Section \ref{secBell}, the S-parameter was  as in (\ref{eq:SparamVartheta})
$$S(\alpha)=3\cos\alpha-\cos(3\alpha)$$
Hence, if the effective state is given by Eq. \eqref{MRHO}, we get
\begin{equation}
S^{\epsilon}(\alpha)=(1-\epsilon) \left(3\cos\alpha-\cos(3\alpha)\right)+\epsilon\left(  2\cos^3\alpha-2\sin^2\alpha \cos(3\alpha)  \right).
\label{Sepsilon}\end{equation}
To control the time delay $T$ in our experiment, we change the length difference between the optical paths in the interferometers (Fig. 1 of main text). We can express $\epsilon$ as a function of the length difference $\Delta L$:
 $$ \epsilon(T) 
= 1 -e^{-T^2 \sigma^2_\omega/2}=1 -e^{-(\Delta L)^2/l_c^2}.$$
When $\Delta L = 0$ the system is in the optical contact condition, and we have $\epsilon = 0$. As we move away from optical contact, increasing $\Delta L$, the parameter $\epsilon$ increases and the coherence of the system is gradually lost.\\
Moreover, in order to take into account the non ideality of the setup
and different sources of noise that contribute to reduce the
visibility of the detection channels, we replace the state
$\rho^\epsilon$ with an effective state $\rho^{eff}$ obtained as a
convex combination of $\rho^\epsilon$ and the maximally mixed state:
$$\rho^{eff}=\eta\,  \rho ^{\epsilon}+(1-\eta)\frac{I}{4},$$
where $\eta\in [0,1]$ is a positive parameter related to the visibility.
The corresponding value for the function $S$ is $$S^{eff}(\alpha)=\eta
S^\epsilon (\alpha),$$ with $S^\epsilon$  given by \eqref{Sepsilon}.

\bibliography{prabell}

\end{document}